\documentclass[11pt]{article}

\usepackage[a4paper,left=1.8cm,right=1.8cm,top=2.0cm,bottom=2.0cm]{geometry}
\usepackage{amsmath,amssymb,amsfonts}
\usepackage{graphicx}
\usepackage{xcolor}
\usepackage{tikz}
\usepackage{cite}
\usepackage[colorlinks=true,linkcolor=blue,citecolor=blue,urlcolor=blue]{hyperref}
\usetikzlibrary{arrows.meta,calc}

\usepackage{mathrsfs}

\definecolor{KBlue}{RGB}{0,55,140}
\definecolor{LGreen}{RGB}{0,105,35}
\definecolor{LandauDeepGreen}{RGB}{18,92,70}
\definecolor{AsymptoticOrange}{RGB}{176,72,20}
\definecolor{MGray}{RGB}{85,85,85}
\definecolor{KFill}{RGB}{247,250,255}
\definecolor{LFill}{RGB}{247,252,248}
\definecolor{MFill}{RGB}{250,250,250}

\title{\bf Morse Bridge between Planar Kepler and\\ Hyperbolic Landau Dynamics} 

\author{
Mikhail S. Plyushchay\\[3pt]
{\small\itshape Departamento de F\'{\i}sica, Universidad de Santiago de Chile,}\\
{\small\itshape Avenida V\'ictor Jara 3493, Santiago, Chile}\\[3pt]
{\small\texttt{\textcolor{blue}{mikhail.plyushchay@usach.cl}}}
}

\date{\empty}

\begin{document}

\maketitle

\begin{abstract}
A common Morse Hamiltonian connects the planar Kepler--Coulomb and
hyperbolic Landau systems, thereby linking flat electric dynamics to
curved magnetic dynamics.  We develop this bridge at both the
classical and quantum levels.
A radial Liouville transformation combined with coupling-constant
metamorphosis converts the separated Kepler problem into the Morse
problem, while reduction at fixed horocyclic momentum converts the
hyperbolic Landau problem into the same system.  In this common
normalization, the Coulomb coupling becomes the signed product of the
magnetic field and the conserved horocyclic momentum, and the quantum
Landau reduction acquires the universal half-density shift \(1/4\).
Going beyond the common reduction, we construct a direct classical
orbit map on the regular nonradial sectors, valid for negative, zero
and positive Kepler energies.  It identifies the Landau height with a
scaled inverse Kepler radius, maps ellipses, parabolas and hyperbolas
to magnetic circles, horocycles and hypercycles, and sends Hamilton's
hodograph affinely to the corresponding Landau carrier circle.  The
reduced Kepler symplectic form and the complex structure selected by
the Binet evolution further determine a distinguished Poincar\'e
K\"ahler metric.  Its nondegenerate Binet orbits have constant
geodesic curvature whose magnitude is the inverse Kepler
eccentricity, while the zero-coupling case gives geodesics. 
The direct orbit map identifies this Binet metric with the Poincar\'e
configuration-space metric of the hyperbolic Landau problem.
  At the quantum level,
composition of the two Liouville maps eliminates the intermediate
Morse wavefunction and yields an exact weighted intertwining identity
between the radial Kepler and horocyclic Landau operator pencils.  At
distinguished half-integer magnetic fields, the angular-momentum
grading of a fixed Kepler shell is reorganized into the finite reduced
Landau tower together with its threshold endpoint.  
These exact correspondences hold in the stated reduced sectors despite
the unitary inequivalence of the complete parent systems.\end{abstract}

\section{Introduction and summary}

Hidden correspondences between solvable systems can expose unexpected
relations between different interactions and geometries, especially
when energies, couplings, coordinates, and even evolution parameters
are exchanged.  Here we establish such correspondences between planar
Kepler--Coulomb dynamics and hyperbolic Landau dynamics.  The common
one-dimensional mediator is the Morse system, but the bridge also
leads to direct classical and reduced quantum relations between the
two parent problems.

The separate ingredients have a substantial history.  Space--time
transformations and path-integral methods relate Coulomb systems to
regular solvable problems
\cite{DuruKleinert1979,DuruKleinert1982,DuruMorse1983,
BarutInomataJunker1,BarutInomataJunker2}, while the hyperbolic Landau
problem and its reduction to a Morse equation have been treated by
geometric, path-integral and group-theoretical methods
\cite{ComHou,Comtet,Grosche,NegroDelOlmoRodriguez}.  The purpose here
is to make the parameter exchanges, reduction data and limiting
states explicit in one convention, and then to extract the direct
Kepler--Landau content of the common reduction.

The novelty of the present construction lies not in either
one-dimensional reduction taken separately, but in their combination
into a single classical and quantum dictionary, in the direct orbit
correspondence revealed by this dictionary, and in the resulting
reduced quantum operator-pencil map.  In the general signed reduction,
the two dictionaries combine to give the cross-relation
\begin{equation}
 \gamma=\mathcal Bp_y ,
 \label{signed-cross-relation-intro}
\end{equation}
which expresses the Coulomb coupling as the product of the hyperbolic
magnetic field and the fixed value of the conserved horocyclic
momentum.  Thus the attractive and repulsive Kepler sectors correspond,
respectively, to
\(\mathcal Bp_y>0\) and \(\mathcal Bp_y<0\).

On the attractive bound-state branch, the quantum dictionary
identifies each Kepler shell with an integer Morse family and maps its
zero-angular-momentum endpoint to a threshold resonance.  It also
transforms the angular-momentum grading of a fixed Kepler shell into
the corresponding finite reduced Landau tower.  At the classical
level, the bridge yields the Kepler-conic form of the Landau time
evolution.  Algebraically, it distinguishes the Casimir reduction of
the magnetic isometry algebra from the parameter-shifting Darboux
structure of the Morse system.

\begin{figure}[!t]
\centering
\resizebox{0.98\textwidth}{!}{%
\begin{tikzpicture}[
  >=Latex,
  every node/.style={align=center,text=black},
  mainarrow/.style={
    -{Latex[length=2.6mm,width=1.8mm]},line width=0.9pt
  },
  downarrow/.style={
    -{Latex[length=2.6mm,width=1.8mm]},line width=0.9pt
  },
]
\node[
  draw=KBlue, fill=KFill, line width=0.9pt, rounded corners=5pt,
  minimum width=4.10cm, minimum height=1.52cm, inner sep=4.8pt
] (K) at (-4.95,1.82) {%
  {\bfseries\large Planar Kepler--Coulomb}\\[1.0mm]
  {\normalsize \({\cal H}_{\rm K}(\gamma)\ {\rm in}\ (r,\varphi)\)}\\[1.0mm]
  {\normalsize \(p_\varphi=\ell,\quad r=e^{-X}\)}
};
\node[
  draw=MGray, fill=MFill, line width=0.9pt, rounded corners=5pt,
  minimum width=2.20cm, minimum height=1.52cm, inner sep=4.8pt
] (M) at (0,1.82) {%
  {\bfseries\large Morse}\\[1.7mm]
  {\normalsize \(H_{\rm M}(C,\lambda)\)}
};
\node[
  draw=LGreen, fill=LFill, line width=0.9pt, rounded corners=5pt,
  minimum width=4.10cm, minimum height=1.52cm, inner sep=4.8pt
] (L) at (4.95,1.82) {%
  {\bfseries\large Hyperbolic Landau}\\[1.0mm]
  {\normalsize
    \(\mathcal H_{\rm L}(\mathcal B)\ {\rm in}\ (X,y)\)}\\[1.0mm]
  {\normalsize \(p_y=C>0,\quad v=e^{-X}\)}
};
\draw[mainarrow,KBlue] (K.east) -- (M.west)
  node[midway,above=1.8pt,font=\normalsize] {r-LT};
\draw[mainarrow,LGreen] (L.west) -- (M.east)
  node[midway,above=1.8pt,font=\normalsize] {h-LT};
\coordinate (KtoM) at ($(K.east)!0.5!(M.west)$);
\coordinate (LtoM) at ($(L.west)!0.5!(M.east)$);
\node[
  draw=KBlue, fill=KFill, line width=0.9pt, rounded corners=5pt,
  minimum width=4.28cm, minimum height=1.82cm, inner sep=4.8pt
] (KD) at (-2.60,-2.00) {%
  {\bfseries\normalsize Kepler--Morse dictionary}\\[1.5mm]
  {\normalsize \(C^2=-2E_{\rm K}\)}\\[0.35mm]
  {\normalsize \(\lambda=\gamma/C\)}\\[0.35mm]
  {\normalsize \(E_{\rm M}=-\ell^2\)}
};
\node[
  draw=LGreen, fill=LFill, line width=0.9pt, rounded corners=5pt,
  minimum width=4.28cm, minimum height=1.82cm, inner sep=4.8pt
] (LD) at (2.60,-2.00) {%
  {\bfseries\normalsize Landau--Morse dictionary}\\[1.5mm]
  {\normalsize \(C=p_y>0\)}\\[0.35mm]
  {\normalsize \(\lambda=\mathcal B\)}\\[0.35mm]
  {\normalsize
    \(E_{\rm M}=\mathcal E_{\rm L}-\mathcal B^2-1/4\)}
};
\draw[downarrow,KBlue] (KtoM) -- (KD.north);
\draw[downarrow,LGreen] (LtoM) -- (LD.north);
\end{tikzpicture}%
}
\caption{Schematic representation of the Kepler--Morse--Landau
bridge.  The planar Kepler--Coulomb system on \(\mathbb R^2\), after
polar separation at fixed \(p_\varphi=\ell\), the logarithmic change
\(r=e^{-X}\), and the radial Liouville transformation (r-LT), gives
the Morse Hamiltonian through a genuine coupling-constant
metamorphosis.  The hyperbolic Landau problem on
\(H^2=\{(u,v):v>0\}\), written in horocyclic coordinates
\(y=u\), \(v=e^{-X}\), after reduction at fixed \(p_y=C>0\) in the
displayed attractive orientation \(\mathcal B>0\) and the horocyclic
Liouville normalization (h-LT), gives the same Morse Hamiltonian.
The two lower boxes show the corresponding parameter dictionaries
for the displayed attractive, negative-energy branch.  In the
general signed Landau reduction,
\(C=|p_y|\) and
\(\lambda=\mathcal B\,\operatorname{sgn}(p_y)\), so that
\(\gamma=C\lambda=\mathcal Bp_y\); the corresponding repulsive
sector is not displayed separately.  The lower-right spectral
relation includes the quantum half-density shift \(1/4\);
classically one has
\(E_{\rm M}={\cal H}_{\mathcal B}-\mathcal B^2\).
The notation \(\mathcal H_{\rm L}(\mathcal B)\) is used
schematically for the Landau Hamiltonian
\({\cal H}_{\mathcal B}\).}
\label{fig:kml-bridge}
\end{figure}
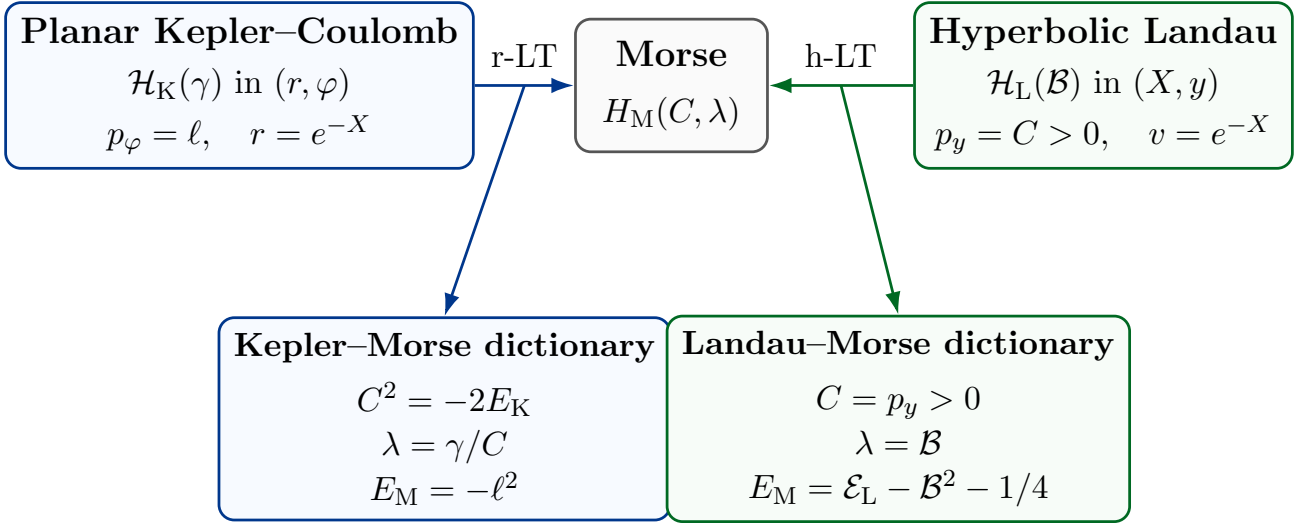

The correspondence is summarized in Fig.~\ref{fig:kml-bridge}.  We use
CCM as an abbreviation for coupling-constant metamorphosis.  See, for
example,
Refs.~\cite{HietarintaCCM,CabralGallasCCM,
HietarintaGrammaticosQuantum,KalMilPost,MillerPostWinternitz}
for CCM and related superintegrable systems.  The bridge is not a
literal identity of the unreduced systems.  Rather, it relates them
by coordinate transformations, reduction at fixed momentum, exchange
of energy and couplings, and, on the Kepler side, a time
reparametrization.  Its physical content is that flat electric
Coulomb dynamics is connected with magnetic dynamics in negative
curvature through a solvable Morse--Whittaker problem.

We normalize the reduced one-dimensional equations so that their
kinetic terms are \(p^2\), or \(-d^2/dX^2\) quantum mechanically.
The Morse Hamiltonian
\cite{Morse,EngQue,Quesne,LemAriAom,DongLemusFrank,CordDab,
CosAleSkaAnd} is
\begin{equation}
 H_{\rm M}(C,\lambda)
 =
 -\frac{d^2}{dX^2}
 +
 V_{\rm M}(X;C,\lambda),
 \qquad
 V_{\rm M}(X;C,\lambda)
 =
 C^2e^{-2X}-2C\lambda e^{-X}.
 \label{HM-intro}
\end{equation}
The Morse potential \(V_{\rm M}\) is a standard example of the
factorization and shape-invariance mechanisms in one-dimensional
quantum mechanics
\cite{InfeldHull,Gendenshtein,CooperKhareSukhatme}.
We take \(C>0\) as the Morse scale parameter.  For \(\lambda>0\),
the minimum of the potential is located at
\begin{equation}
 X_{\min}=\log\frac{C}{\lambda},
 \qquad
 V_{\rm M}(X_{\min})=-\lambda^2.
\end{equation}
Thus, at fixed \(\lambda\), varying \(C\) translates the Morse well
along the \(X\)-axis without changing its depth, whereas \(\lambda\)
determines the depth \(V_{\min}=-\lambda^2\).

On the hyperbolic Landau side, in the displayed attractive orientation,
\begin{equation}
\lambda=\mathcal B>0.
\end{equation}
In terms of the conventional Morse shape parameter \(A_{\rm M}\),
\begin{equation}
\lambda=A_{\rm M}+\frac12.
\end{equation}
On the Kepler side, in the attractive bound-state branch,
\begin{equation}
 C^2=-2E_{\rm K},
 \qquad
 \gamma=C\lambda=C\mathcal B>0,
 \label{KML-dictionary}
\end{equation}
where \(E_{\rm K}<0\) denotes the Kepler energy.  Thus the bridge
parameter \(\lambda=\mathcal B=\gamma/C\) appears as the rescaled
Coulomb strength, the parameter controlling the Morse well depth, and
the hyperbolic magnetic field.

The role of \(C\) is also geometrically meaningful.  Since
\begin{equation}
 V_{\rm M}(X)
 =
 \left(Ce^{-X}-\mathcal B\right)^2-\mathcal B^2,
\end{equation}
changing \(C\) at fixed \(\mathcal B\) translates the position of the
Morse well without changing its depth.  In the attractive branch of
the hyperbolic Landau parent system, the same constant is the fixed
positive horocyclic momentum \(p_y=C>0\), namely the Noether charge
associated with translations along the chosen horizontal horocycles
centered at the ideal point \(\infty\).  This observation will be
useful below and may also point toward an oxidized formulation
\cite{Eisenhart,DuvalBurKPer,Visinescu,CarDuvGibHor}
in which \(C\) is promoted back to a momentum.

The paper is organized as follows.  Sections~2--5 establish the two
Morse reductions and their classical geometric and algebraic content.
Section~6 collects the quantum Morse spectrum, the Kepler shells and
the special magnetic-field values.  Section~7 constructs the direct
classical Kepler--Landau orbit map and its hyperbolic geometry.
Section~8 composes the two quantum Liouville maps into a direct
reduced operator relation and describes the shell-to-tower
intertwiner.  Section~9 discusses the scope of these results and the
remaining global questions.

\section{Kepler--Morse metamorphosis}
We set \(\hbar=1\) and use the kinetic normalization \(m=1/2\).
The Coulomb coupling and the Kepler energy parameter are normalized
so that
\[
 {\cal H}_{\rm K}
 =
 p_r^2+\frac{p_\varphi^2}{r^2}-\frac{2\gamma}{r},
 \qquad
 {\cal H}_{\rm K}=2E_{\rm K}.
\]

After separation of the angular variable, the quantum planar Kepler
problem can be written in one-dimensional radial normal form as
\begin{equation}
 \left[
 -\frac{d^2}{dr^2}
 +
 \frac{\ell^2-\frac14}{r^2}
 -
 \frac{2\gamma}{r}
 \right]u(r)
 =
 2E_{\rm K}u(r),
 \qquad r>0 ,
 \label{radial-Kepler-quantum}
\end{equation}
where \(\ell\in\mathbb Z\) labels the angular mode. The logarithmic
Liouville transformation
\begin{equation}
 r=e^{-X},
 \qquad
 \psi(X)=r^{-1/2}u(r),
 \label{Liouville-r-X}
\end{equation}
has Schwarzian \(\{r,X\}=-1/2\), and hence the Liouville
correction \(-\{r,X\}/2=1/4\). This cancels the \(-1/4\) in the
two-dimensional radial centrifugal term. The transformed equation becomes
\begin{equation}
 \left[
 -\frac{d^2}{dX^2}
 -
 2E_{\rm K}e^{-2X}
 -
 2\gamma e^{-X}
 \right]\psi(X)
 =
 -\ell^2\psi(X).
 \label{Kepler-Morse-Liouville-equation}
\end{equation}
For \(E_{\rm K}<0\), using \(C^2=-2E_{\rm K}\) and
\(\lambda=\gamma/C\), this is precisely
\begin{equation}
 \left[
 -\frac{d^2}{dX^2}
 +
 C^2e^{-2X}
 -
 2C\lambda e^{-X}
 \right]\psi(X)
 =
 E_{\rm M}\psi(X),
 \qquad
 E_{\rm M}=-\ell^2 .
 \label{Morse-from-Kepler}
\end{equation}
Thus the Kepler energy fixes the Morse scale \(C\), the Coulomb
coupling fixes the product \(C\lambda\), 
 and the angular momentum becomes the Morse
spectral parameter.

The classical counterpart is equally transparent. With
\(r=e^{-X}\) and \(p_X=-rp_r\), the classical radial Kepler equation
\(p_r^2+\ell^2/r^2-2\gamma/r=2E_{\rm K}\) becomes
\begin{equation}
 p_X^2-2E_{\rm K}e^{-2X}-2\gamma e^{-X}=-\ell^2 .
 \label{classical-Kepler-Morse}
\end{equation}
The transformation is accompanied by the time reparametrization
\begin{equation}
 dt_{\rm K}=r^2d\tau=e^{-2X}d\tau .
 \label{Kepler-Morse-time-reparam}
\end{equation}
Since \(d\varphi/d\tau=2\ell\), the Kepler polar angle becomes
proportional to the Morse time:
\[
\varphi-\varphi_0=2\ell(\tau-\tau_0).
\]
Consequently the Kepler conic \(r(\varphi)\) becomes an evolution
law for \(r(\tau)=e^{-X(\tau)}\).  The natural Morse variable is the
rescaled coordinate
\[
q=Cr=Ce^{-X}.
\]
For the bound Kepler branch, put
\[
\kappa=|\ell|,
\qquad
e^2
=
1-\frac{\kappa^2}{\lambda^2}
=
1+\frac{2E_{\rm K}\ell^2}{\gamma^2},
\qquad
0\leq e<1.
\]
The resulting evolution law is
\begin{equation}
 q(\tau)
 =
 \frac{\kappa^2}
 {\lambda\left[
 1+e\cos\bigl(2\kappa(\tau-\tau_0)\bigr)
 \right]} .
 \label{Morse-from-Kepler-conic}
\end{equation}
Using \(\lambda=\gamma/C\), this can equivalently be written as
\begin{equation}
 q(\tau)
 =
 C\,\frac{p}
 {1+e\cos\bigl(2\kappa(\tau-\tau_0)\bigr)},
 \qquad
 p=\frac{\ell^2}{\gamma}.
\end{equation}
For \(\ell\neq0\), \(p\) is the usual Kepler semi-latus rectum and,
equivalently, the position of the minimum of the radial effective
potential
\[
V_{\rm eff}(r)=\frac{\ell^2}{r^2}-\frac{2\gamma}{r}.
\]
Hence Eq.~\eqref{Morse-from-Kepler-conic} is not merely analogous to
the Kepler conic equation: after the Liouville transformation it is
precisely the Kepler conic written as the Morse-time evolution of
\(r=e^{-X}\), with
\[
\varphi-\varphi_0=2\ell(\tau-\tau_0);
\]
multiplication by \(C\) gives the natural Morse variable \(q=Cr\).

This is the classical bound motion in the Morse well.  Equivalently,
after shifting the origin of \(\tau\) by half a period, one may write
the denominator of Eq.~\eqref{Morse-from-Kepler-conic} as
\[
\lambda\left[
1-e\cos\bigl(2\kappa(\tau-\tau_0)\bigr)
\right].
\]
In this phase convention the limit \(\kappa\to0\), for which
\(e\to1\), gives the Morse separatrix, while the analytic continuation
\(\kappa\mapsto ik\) gives the positive-energy Morse scattering
trajectory.  This is one precise sense in which Kepler
configuration-space conics reappear as time-dependent Morse
trajectories.

\section{Hyperbolic Landau reduction}
We denote by \(H^2\) the two-dimensional hyperbolic plane, equivalently
Euclidean \(AdS_2\). In Poincar\'e upper-half-plane coordinates,
\begin{equation}
 H^2=\{(u,v):v>0\},
 \qquad
 ds^2=\frac{du^2+dv^2}{v^2}.
 \label{H2-upper-half-plane}
\end{equation}
The homogeneous magnetic field is proportional to the hyperbolic area
form,
\begin{equation}
 F_{\mathcal B}=
 \mathcal B\,\frac{du\wedge dv}{v^2}.
 \label{magnetic-field-H2}
\end{equation}
Introduce horocyclic coordinates
\[
u=y,\qquad v=e^{-X}.
\]
Then \(ds^2=dX^2+e^{2X}dy^2\). In the gauge
\({\cal A}_{\mathcal B}=\mathcal B e^Xdy\), the classical Landau
Hamiltonian 
\cite{ComHou,Comtet,LopAlmTej,Grosche,Beardon,BarrosEtAl,Nair1,Nair2}
in the mass convention \(m=1/2\) is
\begin{equation}
 {\cal H}_{\mathcal B}
 =
 p_X^2+\left(e^{-X}p_y-\mathcal B\right)^2.
 \label{Landau-Hamiltonian-horocyclic}
\end{equation}
The horocyclic coordinate \(y\) is cyclic.  For a general fixed
nonzero value
\begin{equation}
T:=p_y,
\qquad
C=|T|>0,
\qquad
\sigma=\operatorname{sgn}(T),
\end{equation}
the reduced Hamiltonian takes the form
\[
{\cal H}^{\rm red}_{\mathcal B,T}
=
p_X^2+C^2e^{-2X}
-2C(\sigma\mathcal B)e^{-X}
+\mathcal B^2.
\]
Thus the corresponding Morse parameters are
\[
 C=|p_y|,
 \qquad
 \lambda=\mathcal B\,\operatorname{sgn}(p_y).
\]
The attractive Morse branch requires
\(\mathcal Bp_y>0\).  The opposite-sign sector remains part of the
full Landau problem, but after reduction it corresponds classically
to a repulsive Morse potential, which has no potential well and hence
no bounded reduced motion.  Quantum mechanically, the corresponding
fixed reduced sector has no normalizable discrete levels.
 In what follows we choose the magnetic orientation
\(\mathcal B>0\) and consequently restrict the attractive branch to
\(p_y>0\).  We may therefore set
\(p_y=C\), and obtain
\begin{equation}
 {\cal H}^{\rm red}_{\mathcal B,C}
 =
 p_X^2+\left(Ce^{-X}-\mathcal B\right)^2
 =
 H_{\rm M}+\mathcal B^2 .
 \label{Landau-reduced}
\end{equation}
Hence the same Morse Hamiltonian appears with \(\lambda=\mathcal B\).

Quantum mechanically the reduced Landau operator contains a hidden
Liouville step. In horocyclic coordinates,
\(\Delta_{H^2}=\partial_X^2+\partial_X+e^{-2X}\partial_y^2\). After
fixing \(-i\partial_y=C\), the operator is naturally symmetric in
\(e^XdX\). The half-density transformation
\(\Phi=e^{-X/2}\psi\) gives
\[
e^{X/2}\left(-\partial_X^2-\partial_X\right)e^{-X/2}
 =
-\frac{d^2}{dX^2}+\frac14 .
\]
Therefore
\begin{equation}
 \left[
 -\frac{d^2}{dX^2}
 +
 C^2e^{-2X}
 -
 2C\mathcal B e^{-X}
 \right]\psi
 =
 \left(
 {\cal E}_{\rm L}-\mathcal B^2-\frac14
 \right)\psi ,
 \label{quantum-Landau-Morse}
\end{equation}
or
\begin{equation}
 E_{\rm M}={\cal E}_{\rm L}-\mathcal B^2-\frac14 .
 \label{Morse-Landau-energy-quantum}
\end{equation}
The \(1/4\) is the purely quantum horocyclic half-density correction, analogous to the
radial \(-1/4\) correction in two-dimensional Euclidean central
problems.

\section{Classical hyperbolic Landau orbits and magnetic
\texorpdfstring{\(SL(2,\mathbb R)\)}{SL(2,R)} symmetry}
The orientation-preserving isometry group of \(H^2\) is
\(PSL(2,\mathbb R)\).  Below we use \(SL(2,\mathbb R)\) for its
double cover and \(sl(2,\mathbb R)\) for the corresponding Lie
algebra.  Since the two-form \(F_{\mathcal B}\) is
\(\mathcal B\) times the hyperbolic area form, it is invariant under
this group.  The magnetic Noether charges may be chosen as
\[
 J_-=p_y,\qquad J_0=yp_y-p_X,
\]
and
\[
 J_+=(y^2-e^{-2X})p_y-2yp_X+2\mathcal B e^{-X}.
\]
This is the parabolic, or Iwasawa, basis adapted to horocyclic
reduction.  For \(\mathcal B=0\), \(J_-\) generates translations,
whereas \(J_0\) is the noncompact split Cartan generator of
dilations:
\begin{equation}
 X_{J_-}=\partial_u,
 \qquad
 X_{J_0}=u\partial_u+v\partial_v.
\end{equation}
In particular, the dilation flow generated by \(J_0\) does not fix
the point \(z=i\).  The compact \(SO(2)\) generator stabilizing this
point is instead the elliptic combination
\begin{equation}
 R=\frac12\left(J_-+J_+\right).
\end{equation}
Indeed, for \(\mathcal B=0\), its Killing vector field is
\begin{equation}
 X_R
 =
 \frac12\left(1+u^2-v^2\right)\partial_u
 +
 uv\partial_v,
 \qquad
 X_R\big|_{z=i}=0.
\end{equation}
For \(\mathcal B\neq0\), the Noether charges acquire the corresponding
magnetic corrections and close into \(sl(2,\mathbb R)\), while their
projected coordinate-space Killing fields generate the same
\(SL(2,\mathbb R)\) isometries.  In particular, the magnetically
corrected combination \(\frac12(J_-+J_+)\) generates the compact
stabilizer of \(z=i\).  Their Casimir satisfies
\begin{equation}
 {\cal C}_{\rm L}=J_0^2-J_-J_+
 =
 {\cal H}_{\mathcal B}-\mathcal B^2 .
 \label{Landau-Casimir}
\end{equation}
Fixing \(J_-=C\), the trajectory in the upper-half-plane coordinates
\((u,v)=(y,e^{-X})\) is
\begin{equation}
 \left(u-\frac{J_0}{C}\right)^2
 +
 \left(v-\frac{\mathcal B}{C}\right)^2
 =
 \frac{{\cal H}_{\mathcal B}}{C^2}.
 \label{Landau-trajectory-circle}
\end{equation}
Thus, in the upper-half-plane model, Landau trajectories are represented
by Euclidean circles, or by their arcs lying in \(v>0\).
When the complete Euclidean circle lies inside the upper half-plane,
is tangent to the boundary \(v=0\), or intersects it, the corresponding
hyperbolic trajectory is respectively a closed magnetic circle, a
horocycle, or an open hypercycle:
\[
{\cal H}_{\mathcal B}<\mathcal B^2,\qquad
{\cal H}_{\mathcal B}=\mathcal B^2,\qquad
{\cal H}_{\mathcal B}>\mathcal B^2 .
\]
Equivalently, since \(E_{\rm M}={\cal H}_{\mathcal B}-\mathcal B^2\)
classically, the signs \(E_{\rm M}<0\), \(E_{\rm M}=0\), and
\(E_{\rm M}>0\) correspond respectively to closed magnetic circles,
horocycles, and open hypercycles. In the Beltrami--Klein disk these
same trajectories become projective conics classified by their
relation to the absolute: the circle-type orbits do not meet the
absolute, horocycles are tangent to it, and hypercycles meet it at two
ideal points. This parallels the Kepler ellipse/parabola/hyperbola
classification relative to infinity.

The circle representation \eqref{Landau-trajectory-circle} assumes
\(C=p_u\neq0\) and therefore does not directly include the horizontal
horocycle.  The latter is the horocycle centered at the ideal point
\(u=\infty\).  It can be obtained as a correlated limit of the
finite horocycles
\[
 (u-u_0)^2+(v-R)^2=R^2,
 \qquad
 u_0=\frac{J_0}{C},
 \qquad
 R=\frac{\mathcal B}{C},
 \qquad
 {\cal H}_{\mathcal B}=\mathcal B^2,
\]
in which
\[
 u_0\longrightarrow\infty,
 \qquad
 R\longrightarrow\infty,
 \qquad
 \frac{u_0^2}{2R}\longrightarrow v_*>0.
\]
Indeed, after division of the circle equation by \(2R\), this limit
gives
\[
 v=v_*.
\]
Dynamically, it belongs to the limiting sector \(C=p_u=0\):
for \(p_v=0\) and \({\cal H}_{\mathcal B}=\mathcal B^2\),
Hamilton's equations give
\[
 v(t)=v_*,
 \qquad
 \dot u=-2\mathcal Bv_*.
\]
Thus the particle moves along a horizontal horocycle.  Equivalently,
this curve is related by a M\"obius isometry of \(H^2\) to any
horocycle tangent to \(v=0\) at a finite ideal point.

\section{Landau time evolution and conic law}
There is also a dynamical form of the same correspondence. In the
upper-half-plane variables the Hamiltonian may be written as
\[
{\cal H}_{\mathcal B}
 =
v^2p_v^2+\left(vT-\mathcal B\right)^2,
\qquad
T=p_u=p_y .
\]
For \(x=1/v\), Hamilton's equations give the linear equation
\begin{equation}
 \ddot x
 =
4\left({\cal H}_{\mathcal B}-\mathcal B^2\right)x
+
4\mathcal B T .
 \label{Landau-x-linear-equation}
\end{equation}
The sign condition
\[
\mathcal B T>0
\]
selects the attractive Morse branch of the reduced system.  This
condition should be distinguished from the condition for bounded
motion.  Since the reduced Morse energy is
\[
E_{\rm M}
=
{\cal H}_{\mathcal B}-\mathcal B^2,
\]
the attractive branch contains three dynamical regimes:
\[
E_{\rm M}<0,\qquad E_{\rm M}=0,\qquad E_{\rm M}>0,
\]
corresponding respectively to bound Morse motion, the threshold
separatrix, and scattering motion.  On the hyperbolic Landau side
these are precisely the closed magnetic-circle, horocycle, and
hypercycle regimes.  In particular, closed motion requires
\[
{\cal H}_{\mathcal B}<\mathcal B^2.
\]
Notice that this inequality automatically implies
\(\mathcal B T>0\); the converse is not true.

In the closed-orbit regime, put
\[
\kappa^2
=
\mathcal B^2-{\cal H}_{\mathcal B}>0,
\qquad
\omega=2\kappa.
\]
The conserved Hamiltonian fixes the oscillation amplitude, and the
solution can be written as
\begin{equation}
 v(t)
 =
 \frac{p}{1+e\cos\omega(t-t_0)},
 \qquad
 p=\frac{\kappa^2}{\mathcal B T},
 \qquad
 e=\frac{\sqrt{{\cal H}_{\mathcal B}}}{|\mathcal B|}<1 .
 \label{Landau-v-kepler-form}
\end{equation}
This is the Kepler conic law with the Kepler polar angle replaced by
the rescaled Landau time \(\theta=\omega(t-t_0)\).
The sign of the cosine is conventional and can be changed
by shifting \(t_0\) by half a period. The parabolic and hyperbolic
analogues are obtained at \({\cal H}_{\mathcal B}=\mathcal B^2\) and
by the continuation \(\kappa\mapsto ik\).

More intrinsically, the three dynamical regimes are distinguished by
the sign of
\[
\Delta_{\mathcal B}
:=
{\cal H}_{\mathcal B}-\mathcal B^{2}.
\]
Indeed,
\[
e^{2}-1
=
\frac{{\cal H}_{\mathcal B}-\mathcal B^{2}}{\mathcal B^{2}},
\]
so that
\[
\Delta_{\mathcal B}<0,\qquad
\Delta_{\mathcal B}=0,\qquad
\Delta_{\mathcal B}>0
\]
correspond respectively to
\[
e<1,\qquad e=1,\qquad e>1.
\]
This is the direct Landau analogue of the Kepler classification
\(E_{\rm K}<0\), \(E_{\rm K}=0\), and \(E_{\rm K}>0\) into elliptic,
parabolic, and hyperbolic trajectories.  On the Landau side, the same
three cases describe a closed magnetic circle, a horocycle, and an
open hypercycle.  For
\[
k^{2}={\cal H}_{\mathcal B}-\mathcal B^{2}>0,
\]
the phase-shifted trigonometric solution analytically continues under
\(\kappa\mapsto ik\) to
\[
v(t)=
\frac{k^{2}}
{T\left[
\sqrt{{\cal H}_{\mathcal B}}\,
\cosh 2k(t-t_{0})-\mathcal B
\right]}.
\]
The hypercycle has two endpoints on the ideal boundary \(v=0\), which
are approached asymptotically as \(t\to\pm\infty\).

After fixing \(T=C\), multiplying by \(T\) gives
\(q=Tv=Ce^{-X}\), and hence exactly the Morse evolution laws:
\begin{equation}
q(t)=
\begin{cases}
\dfrac{\kappa^2}
{\mathcal B+\sqrt{{\cal H}_{\mathcal B}}\,
 \cos 2\kappa(t-t_0)},
&
{\cal H}_{\mathcal B}<\mathcal B^2,
\\[4mm]
\dfrac{2\mathcal B}
{1+4\mathcal B^2(t-t_0)^2},
&
{\cal H}_{\mathcal B}=\mathcal B^2,
\\[4mm]
\dfrac{k^2}
{\sqrt{\mathcal B^2+k^2}\,
 \cosh 2k(t-t_0)-\mathcal B},
&
{\cal H}_{\mathcal B}=\mathcal B^2+k^2 .
\end{cases}
\label{Landau-Morse-evolution-cases}
\end{equation}
The first line is written with the Kepler sign convention. The
equivalent phase-shifted form with
\(-\sqrt{{\cal H}_{\mathcal B}}\cos 2\kappa(t-t_0)\) gives the smooth
threshold limit and the analytic continuation to the scattering line.
Thus the Morse time evolution of \(q(t)\) is the reduced form of the
Landau time evolution of \(v(t)\).

The cyclic coordinate \(u\), suppressed in the one-dimensional Morse
reduction, reconstructs the full two-dimensional orbit. Since
\(J_0=up_u+vp_v\), one has
\begin{equation}
 u(t)=\frac{J_0}{T}+\frac{\dot x}{2Tx}.
 \label{u-reconstruction}
\end{equation}
In the closed case, with
\(\theta=2\kappa(t-t_0)\), Eq.~\eqref{u-reconstruction} gives,
up to the choice of phase,
\[
 u-\frac{J_0}{T}
 =
 -\frac{\sqrt{{\cal H}_{\mathcal B}}}{\kappa}\,
 v\sin\theta .
\]
Thus \(u-J_0/T\) is proportional to the transverse Cartesian
coordinate of the associated Kepler conic, while \(v(t)\) obeys its
radial law. The Morse reduction retains this radial/conic law and
suppresses the second coordinate needed to reconstruct the full
two-dimensional orbit. A genuine unreduced Kepler--Landau
correspondence would have to restore this coordinate together with
the corresponding time variable.

\section{Quantum spectrum, Kepler shells and special magnetic field values}
\label{sec:quantum-spectrum}
The Morse bound-state energies are
\begin{equation}
 E_{{\rm M},n}=-(A_{\rm M}-n)^2,
 \qquad
 n=0,1,2,\ldots,\qquad n<A_{\rm M}.
 \label{Morse-bound-spectrum}
\end{equation}
Using the quantum relation
\(E_{\rm M}={\cal E}_{\rm L}-\mathcal B^2-1/4\), this gives
\begin{equation}
 {\cal E}_{{\rm L},n}
 =
 \mathcal B^2+\frac14
 -
 \left(\mathcal B-\frac12-n\right)^2
 =
 \mathcal B(2n+1)-n(n+1),
 \label{Landau-discrete-levels}
\end{equation}
with \(n<\mathcal B-1/2\). The continuum starts at
\begin{equation}
 {\cal E}_{\rm th}=\mathcal B^2+\frac14 .
 \label{Landau-continuum-threshold}
\end{equation}

The same Morse spectrum also has a direct Kepler interpretation.  For a
fixed attractive Coulomb coupling \(\gamma>0\), the two-dimensional
Kepler bound spectrum in the present normalization is
\begin{equation}
 E_{{\rm K},n_r,\ell}
 =
 -\frac{\gamma^2}
 {2\left(n_r+|\ell|+\frac12\right)^2},
 \qquad
 n_r=0,1,\ldots,\qquad \ell\in\mathbb Z .
 \label{Kepler-bound-spectrum-expanded}
\end{equation}
The Kepler--Morse dictionary then gives
\begin{equation}
 C=\sqrt{-2E_{\rm K}}
 =
 \frac{\gamma}{n_r+|\ell|+\frac12},
 \qquad
 \lambda=\frac{\gamma}{C}
 =
 n_r+|\ell|+\frac12,
 \qquad
 A_{\rm M}=n_r+|\ell|.
 \label{Kepler-Morse-quantum-dictionary-expanded}
\end{equation}
Therefore physical Kepler bound states select the integer Morse
families.  Moreover,
\begin{equation}
 E_{\rm M}=-\ell^2
 =
 -\left(A_{\rm M}-n_r\right)^2 ,
 \label{Kepler-Morse-level-identification-expanded}
\end{equation}
so the Morse level number is precisely the Kepler radial quantum
number, \(n=n_r\).
For fixed \(A_{\rm M}=N\), the Kepler shell
\(n_r+|\ell|=N\) is mapped to the finite Morse chain
\(n=0,1,\ldots,N-1\), together with the formal endpoint \(n=N\).
The normalizable Morse states correspond to \(|\ell|=N-n>0\), while
the \(\ell=0\) member of the Kepler shell corresponds, under the
Liouville spectral-equation map, to the Morse threshold resonance
\(n=N\), \(E_{\rm M}=0\).  The sign
degeneracy \(\ell\leftrightarrow-\ell\) is, of course, suppressed by
the radial reduction.

The integer Morse values \(A_{\rm M}=N\) correspond to half-integer
magnetic fields,
\begin{equation}
 \mathcal B=N+\frac12.
 \label{integer-Morse-half-integer-B}
\end{equation}
For these values the highest would-be Landau level reaches the
continuum edge:
\[
{\cal E}_{{\rm L},N}={\cal E}_{\rm th}.
\]
On the Morse side this is the threshold resonance at \(E_{\rm M}=0\):
the state is bounded but not square-integrable. For generic
\(A_{\rm M}\notin\mathbb Z_{\ge0}\), the threshold solution selected by
regularity at the Morse wall grows linearly as \(X\to+\infty\), and no
bounded threshold resonance is present.

The one-channel Morse reflection amplitude is
\begin{equation}
 {\cal R}_{A_{\rm M}}(k)
 =
 (2C)^{-2ik}
 \frac{\Gamma(2ik)}{\Gamma(-2ik)}
 \frac{\Gamma(-A_{\rm M}-ik)}
      {\Gamma(-A_{\rm M}+ik)}.
 \label{Morse-reflection-amplitude}
\end{equation}
For \(A_{\rm M}=N\), it factorizes as
\begin{equation}
 {\cal R}_{N}(k)
 =
 {\cal R}_0(k)
 \prod_{j=1}^{N}\frac{j-ik}{j+ik}.
 \label{integer-Morse-Blaschke}
\end{equation}
Thus the special half-integer magnetic fields are the Landau-side image
of finite Blaschke-type Morse scattering data and of the Morse
threshold resonance.

The same integer pattern is the scattering counterpart of the
Darboux--SUSY factorization of the Morse Hamiltonian: each Darboux step
removes the lowest bound state and shifts the shape parameter by one
unit, while at integer \(A_{\rm M}\) the chain ends at a threshold
resonance rather than at a normalizable state
\cite{InfeldHull,Gendenshtein,CooperKhareSukhatme}.
This factorization is also the quantum trace of the Morse potential
algebra.  In this paragraph we write
\[
A\equiv A_{\rm M},
\qquad
H_A(C)\equiv H_{\rm M}(C,\lambda=A+1/2).
\]
With
\[
 {\cal D}_{A}=\frac{d}{dX}+A-Ce^{-X},
 \qquad
 {\cal D}_{A}^{\dagger}=-\frac{d}{dX}+A-Ce^{-X},
\]
one has
\[
 {\cal D}_{A}^{\dagger}{\cal D}_{A}=H_A(C)+A^2,
 \qquad
 {\cal D}_{A}{\cal D}_{A}^{\dagger}=H_{A-1}(C)+A^2 .
\]
The Darboux step therefore shifts the conventional Morse parameter
\(A=A_{\rm M}\) as \(A_{\rm M}\mapsto A_{\rm M}-1\), or equivalently
\(\lambda\mapsto\lambda-1\) and \(\mathcal B\mapsto\mathcal B-1\).
In an oxidized description, where \(A_{\rm M}\) is represented by a
momentum, these intertwiners become ladder operators of a noncompact
spectrum-generating, or potential, algebra.

\section{Direct classical Kepler--Landau correspondence}
\label{sec:direct-KL}

The Kepler--Morse and Landau--Morse reductions established above encode
the radial Kepler equation at fixed angular momentum \(\ell\) and the
horocyclic Landau equation at fixed momentum \(T=p_u\) in the same
Morse Hamiltonian. The parameter and conic dictionaries generated by
this common Morse bridge suggest that the classical orbits of the
Kepler and Landau systems should admit a direct correspondence. We now
show that this expectation is indeed realized: the correspondence can
be formulated explicitly at the level of the classical orbit data,
without the need to pass through a real intermediate Morse trajectory.
We first construct the resulting map on the regular sector
\[
\gamma>0,\qquad \ell\neq0,\qquad T=p_u\neq0.
\]
Changing the sign of \(\ell\) reverses the Kepler orientation, while
the simultaneous reversal
\((\mathcal B,T)\mapsto(-\mathcal B,-T)\) preserves the attractive
Landau branch and reverses its orientation.  
Reversing only \(\mathcal B\) or only \(T\) changes the sign of
\(\mathcal BT\) and therefore passes, on the Landau side, to the
repulsive Morse branch; its direct Kepler counterpart has
\(\gamma<0\), as discussed below.

\subsection{Common Binet system and the scaled orbit map}
Put \(q=1/r\) and use the Kepler polar angle
\(\alpha=\varphi-\varphi_0\) as the orbit parameter.  A prime denotes
\(d/d\alpha\).  Hamilton's equations imply
\(q'=-p_r/\ell\) and yield the Binet equation, while the conserved
Kepler energy \(E_{\rm K}\) fixes its first integral:
\begin{equation}
 q''+q=\frac{\gamma}{\ell^2},
 \qquad
 (q')^2+
 \left(q-\frac{\gamma}{\ell^2}\right)^2
 =
 e^2\frac{\gamma^2}{\ell^4},
 \qquad
 e^2
 =
 1+\frac{2E_{\rm K}\ell^2}{\gamma^2}.
 \label{Kepler-Binet-direct}
\end{equation}

On the Landau side let
\begin{equation}
 u_0=\frac{J_0}{T}=u+\frac{vp_v}{T},
 \qquad
 U=u-u_0,
 \qquad
 \Lambda=\frac{\mathcal B}{T}>0.
 \label{Landau-center-direct}
\end{equation}
Equation~\eqref{Landau-trajectory-circle} becomes
\begin{equation}
 U^2+(v-\Lambda)^2=(e\Lambda)^2,
 \qquad
 e^2=\frac{{\cal H}_{\mathcal B}}{\mathcal B^2}.
 \label{Landau-circle-direct}
\end{equation}
It admits the common angular parametrization
\begin{equation}
 U=e\Lambda\sin\alpha,
 \qquad
 v=\Lambda(1+e\cos\alpha).
 \label{Landau-angle-direct}
\end{equation}
Equations~\eqref{Kepler-Binet-direct} and
\eqref{Landau-circle-direct} describe Euclidean circles with the same
dimensionless eccentricity \(e\), but with generally different
centers,
\[
q_0=\frac{\gamma}{\ell^2},
\qquad
\Lambda=\frac{\mathcal B}{T}.
\]
Their identification therefore requires a positive homothety factor,
which we define by
\[
a:=\frac{\Lambda}{q_0}
=\frac{\mathcal B\ell^2}{T\gamma}>0.
\]
Equivalently,
\[
\frac{\mathcal B}{T}
=a\frac{\gamma}{\ell^2}.
\]
The factor \(a\) is not an additional physical parameter. It records
the relative normalization of the Kepler inverse-radius coordinate
\(q=1/r\) and the upper-half-plane Landau coordinates; it also supplies
the corresponding conversion of units if dimensional coordinates are
used. In terms of the two Binet planes, the map is the scaled
reflection
\[
(U,v)=a(-q',q).
\]
Using \(q=1/r\), \(q'=-p_r/\ell\), and
\(U=-vp_v/T\), this gives
\begin{equation}
 v=\frac{a}{r},
 \qquad
 U=\frac{a}{\ell}p_r,
 \qquad
 p_v=-\frac{Tr}{\ell}p_r,
 \qquad
 \frac{\mathcal B}{T}
 =a\frac{\gamma}{\ell^2}.
 \label{scaled-direct-map}
\end{equation}
The third equality follows from
\(U=-vp_v/T\).  Substitution into the Landau Hamiltonian gives the
exact identity
\begin{equation}
 {\cal H}_{\mathcal B}-\mathcal B^2
 =\frac{a^2T^2}{\ell^2}{\cal H}_{\rm K}
 =\frac{2a^2T^2}{\ell^2}E_{\rm K}.
 \label{direct-energy-map}
\end{equation}
Consequently,
\begin{equation}
 e_{\rm K}^2
 =1+\frac{2E_{\rm K}\ell^2}{\gamma^2}
 =\frac{{\cal H}_{\mathcal B}}{\mathcal B^2}
 =e_{\rm L}^2.
 \label{direct-eccentricity-map}
\end{equation}
Thus the change \(E_{\rm K}<0\to E_{\rm K}>0\) is represented
directly by
\({\cal H}_{\mathcal B}-\mathcal B^2<0\to
{\cal H}_{\mathcal B}-\mathcal B^2>0\), without using the analytic
continuation \(\kappa\mapsto ik\) as the definition of the open orbit.
For
\(e>1\), only the arc of the carrier circle lying in \(v>0\) is
physical; its two ideal endpoints occur at a finite interval of
\(\alpha\), but are reached only as \(t_{\rm L}\to\pm\infty\).

The orbit angles agree up to an orientation and an additive constant.
The physical times do not.  Hamilton's equations applied to
Eq.~\eqref{scaled-direct-map} give
\begin{equation}
 dt_{\rm L}=\frac{\ell}{aTr}\,dt_{\rm K},
 \qquad
 \frac{d\alpha}{dt_{\rm L}}=2Tv,
 \qquad
 \frac{d\varphi}{dt_{\rm K}}=\frac{2\ell}{r^2}.
 \label{direct-time-map}
\end{equation}
With \(u=u_0+ap_r/\ell\) and the expression for \(p_v\) in
Eq.~\eqref{scaled-direct-map}, the remaining Landau Hamilton equations
follow from the Kepler equations under the same time change.
This reparametrization explains why the same finite orbit-angle
interval can correspond to infinite Landau time.  The map is an exact
map of reduced orbit data and their parametrizations.  Since it uses
the fixed constants \(\ell,T\) and the orbit center \(u_0\), it is not
by itself a symplectomorphism of the two unreduced four-dimensional
phase spaces.

For reference, the principal entries of the direct dictionary are
collected in Table~\ref{tab:direct-KL}.

\begin{table}[t]
\centering
\small
\resizebox{\textwidth}{!}{%
\begin{tabular}{|c|c|c|}
\hline
Planar Kepler & Hyperbolic Landau & Meaning \\
\hline
\(q=1/r\) & \(v/a\) & Binet radial variable \\
\(-p_r/\ell=q'\) & \(vp_v/(aT)=-U/a\) & derivative with respect to the common orbit angle \\
\(\gamma/\ell^2\) & \(\mathcal B/(aT)\) & center of the Binet circle \\
\(2E_{\rm K}\) & \(\ell^2({\cal H}_{\mathcal B}-\mathcal B^2)/(a^2T^2)\) & signed reduced energy \\
\(e_{\rm K}\) & \(e_{\rm L}=\sqrt{{\cal H}_{\mathcal B}}/|\mathcal B|\) & common eccentricity \\
ellipse / parabola / hyperbola & magnetic circle / horocycle / hypercycle & \(e<1\), \(e=1\), \(e>1\) \\
\(\varphi-\varphi_0\) & carrier-circle angle \(\alpha\) & common orbit parameter \\
\(dt_{\rm K}\) & \(dt_{\rm L}=\ell\,dt_{\rm K}/(aTr)\) & physical-time reparametrization \\
\hline
\end{tabular}%
}
\caption{The direct classical Kepler--Landau dictionary on a regular
sector.  Here \(a=\mathcal B\ell^2/(T\gamma)\) is the positive homothety
factor matching the two Binet circles; it reflects their relative
coordinate normalization rather than an additional physical
parameter.}
\label{tab:direct-KL}
\end{table}

\subsection{Kepler hodographs and Landau carrier circles}

The direct relation has a particularly transparent momentum-space
interpretation.  Choose the positive second axis in momentum space along the Kepler
momentum at periapsis and put \(\rho=\gamma/\ell\), with
\(\ell>0\).  Hamilton's hodograph \cite{HamiltonHodograph,GibbonsHodograph}
is
\begin{equation}
 p_\perp=-\rho\sin\alpha,
 \qquad
 p_\parallel=\rho(e+\cos\alpha),
 \qquad
 p_\perp^2+(p_\parallel-e\rho)^2=\rho^2.
 \label{Kepler-hodograph-circle}
\end{equation}
For \(e<1\), \(e=1\) and \(e>1\), the momentum origin lies inside,
on, or outside this circle.  In the last case the physical Kepler
hodograph is only the arc corresponding to \(r>0\).

For \(e>0\), Eqs.~\eqref{Landau-angle-direct} and
\eqref{Kepler-hodograph-circle} are related by the affine map
\begin{equation}
 U=-\frac{\Lambda e}{\rho}p_\perp,
 \qquad
 v=\Lambda(1-e^2)+\frac{\Lambda e}{\rho}p_\parallel.
 \label{hodograph-Landau-affine}
\end{equation}
This is a one-to-one map from the Kepler hodograph to the corresponding
Landau carrier circle, restricted to the physical arcs when \(e>1\).
It realizes explicitly the momentum--configuration exchange suggested
by the two pictures: a Kepler momentum coordinate becomes a Landau
configuration coordinate.  The statement is orbitwise, however,
because its scale, translation and distinguished direction depend on
\(E_{\rm K},\ell,\gamma\) and on the Laplace--Runge--Lenz vector.

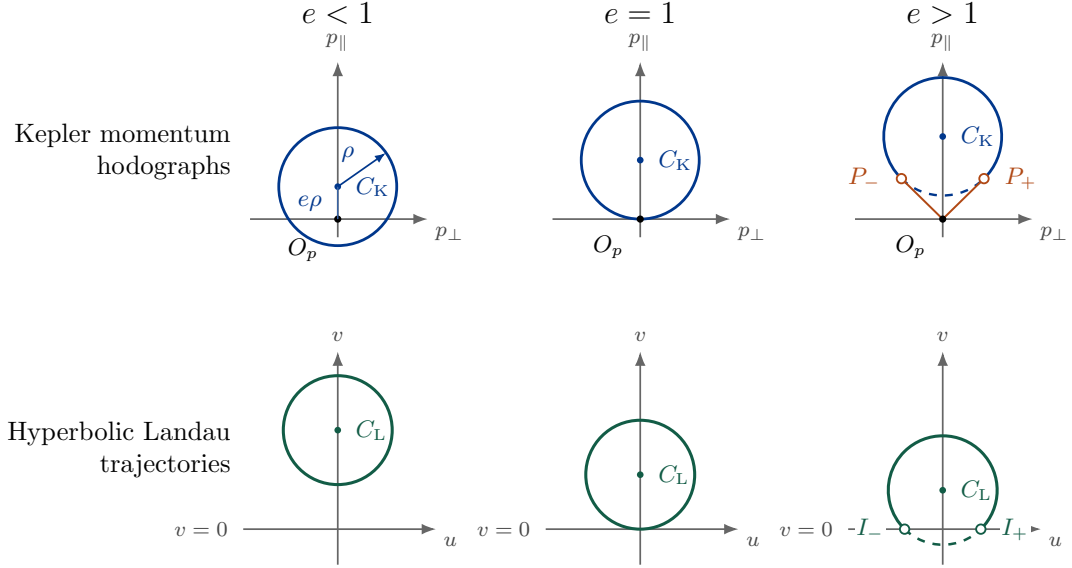
\begin{figure}[t]
\centering
\begin{tikzpicture}[
  x=1cm,y=1cm,>=Latex,
  axis/.style={black!60,line width=0.65pt,-{Latex[length=1.9mm]}},
  axislabel/.style={font=\scriptsize,text=black!72},
  kcurve/.style={KBlue,line width=1.05pt},
  kdash/.style={KBlue,line width=0.9pt,dashed},
  lcurve/.style={LandauDeepGreen,line width=1.15pt},
  ldash/.style={LandauDeepGreen,line width=0.95pt,dashed},
  tangent/.style={AsymptoticOrange,line width=0.75pt},
  endpoint/.style={circle,draw,fill=white,inner sep=1.25pt,line width=0.7pt},
  klabel/.style={font=\footnotesize,text=KBlue},
  llabel/.style={font=\footnotesize,text=LandauDeepGreen},
  plabel/.style={font=\footnotesize,text=AsymptoticOrange}
]
  \def\R{0.78}
  \def\eone{0.55}
  \def\ehyp{1.40}
  \def\yO{4.70}
  \def\yB{0.60}
  \def\LR{0.72}
  \pgfmathsetmacro{\beta}{asin(1/\ehyp)}

  \foreach \x/\lab in {0/{\(e<1\)},4.0/{\(e=1\)},8.0/{\(e>1\)}}
    {\node[font=\large] at (\x,7.43) {\lab};}

  \node[anchor=east,align=right,font=\small] at (-1.28,5.60)
    {Kepler momentum\\hodographs};
  \node[anchor=east,align=right,font=\small] at (-1.28,1.66)
    {Hyperbolic Landau\\trajectories};

  \foreach \x in {0,4.0,8.0}{
    \draw[axis] (\x-1.16,\yO)--(\x+1.19,\yO)
      node[below right=-1pt,axislabel] {\(p_\perp\)};
    \draw[axis] (\x,\yO-0.24)--(\x,6.78)
      node[above,axislabel] {\(p_\parallel\)};
  }

  \coordinate (KO1) at (0,\yO);
  \coordinate (KC1) at (0,{\yO+\eone*\R});
  \draw[kcurve] (KC1) circle (\R);
  \fill[black] (KO1) circle (1.35pt)
    node[below left=3.5pt,font=\footnotesize] {\(O_p\)};
  \fill[KBlue] (KC1) circle (1.25pt)
    node[right=3pt,klabel] {\(C_{\rm K}\)};
  \draw[KBlue!78,line width=0.55pt] (KO1)--(KC1)
    node[midway,left=2pt,klabel] {\(e\rho\)};
  \draw[KBlue,line width=0.65pt,-{Latex[length=1.5mm]}]
    (KC1)--($(KC1)+(35:\R)$)
    node[midway,above left=-1pt,klabel] {\(\rho\)};

  \coordinate (KO2) at (4.0,\yO);
  \coordinate (KC2) at (4.0,{\yO+\R});
  \draw[kcurve] (KC2) circle (\R);
  \fill[black] (KO2) circle (1.35pt)
    node[below left=2.5pt,font=\footnotesize] {\(O_p\)};
  \fill[KBlue] (KC2) circle (1.25pt)
    node[right=3pt,klabel] {\(C_{\rm K}\)};

  \coordinate (KO3) at (8.0,\yO);
  \coordinate (KC3) at (8.0,{\yO+\ehyp*\R});
  \coordinate (KPp) at ($(KC3)+(-\beta:\R)$);
  \coordinate (KPm) at ($(KC3)+({180+\beta}:\R)$);
  \draw[kcurve] (KPp)
    arc[start angle=-\beta,end angle={180+\beta},radius=\R];
  \draw[kdash] (KPm)
    arc[start angle={180+\beta},end angle={360-\beta},radius=\R];
  \draw[tangent] (KO3)--(KPp);
  \draw[tangent] (KO3)--(KPm);
  \node[endpoint,draw=AsymptoticOrange] at (KPp) {};
  \node[endpoint,draw=AsymptoticOrange] at (KPm) {};
  \node[plabel,right=4pt] at (KPp) {\(P_+\)};
  \node[plabel,left=4pt] at (KPm) {\(P_-\)};
  \fill[black] (KO3) circle (1.35pt)
    node[below left=2.5pt,font=\footnotesize] {\(O_p\)};
  \fill[KBlue] (KC3) circle (1.25pt)
    node[right=3pt,klabel] {\(C_{\rm K}\)};

  \foreach \x in {0,4.0,8.0}{
    \draw[axis] (\x-1.25,\yB)--(\x+1.30,\yB)
      node[below right=-1pt,axislabel] {\(u\)};
    \draw[axis] (\x,{\yB-0.47})--(\x,2.95)
      node[above,axislabel] {\(v\)};
    \node[anchor=east,axislabel] at (\x-1.32,\yB) {\(v=0\)};
  }


  \coordinate (LC1) at (0,{\yB+\LR/\eone});
  \draw[lcurve] (LC1) circle (\LR);
  \fill[LandauDeepGreen] (LC1) circle (1.25pt)
    node[right=3pt,llabel] {\(C_{\rm L}\)};

  \coordinate (LC2) at (4.0,{\yB+\LR});
  \draw[lcurve] (LC2) circle (\LR);
  \fill[LandauDeepGreen] (LC2) circle (1.25pt)
    node[right=3pt,llabel] {\(C_{\rm L}\)};

  \coordinate (LC3) at (8.0,{\yB+\LR/\ehyp});
  \coordinate (LIp) at ($(LC3)+(-\beta:\LR)$);
  \coordinate (LIm) at ($(LC3)+({180+\beta}:\LR)$);
  \draw[lcurve] (LIp)
    arc[start angle=-\beta,end angle={180+\beta},radius=\LR];
  \draw[ldash] (LIm)
    arc[start angle={180+\beta},end angle={360-\beta},radius=\LR];
  \node[endpoint,draw=LandauDeepGreen] at (LIp) {};
  \node[endpoint,draw=LandauDeepGreen] at (LIm) {};
  \node[
    llabel,
    right=7pt,
    fill=white,
    inner sep=0.6pt
  ] at (LIp) {\(I_+\)};
  \node[
    llabel,
    left=7pt,
    fill=white,
    inner sep=0.6pt
  ] at (LIm) {\(I_-\)};
  \fill[LandauDeepGreen] (LC3) circle (1.25pt)
    node[right=3pt,llabel] {\(C_{\rm L}\)};
\end{tikzpicture}
\caption{Kepler momentum hodographs (upper row) and the corresponding
Landau carrier circles in the upper-half-plane (lower row).  In the
upper row, \(O_p\) is the momentum origin, \(C_{\rm K}\) is the
hodograph center, the circle radius is \(\rho\), and
\(\lvert O_pC_{\rm K}\rvert=e\rho\) is the distance from the momentum
origin to the center.  For \(e>1\), the physical hodograph is the
solid major arc.  The rays
\(O_pP_\pm\) are tangent to the hodograph circle and give the two
asymptotic velocity directions; these are parallel to the asymptotes
of the Kepler hyperbola in configuration space.  In the lower row,
the horizontal axis is the ideal boundary \(v=0\), the Landau carrier
circle has center \(C_{\rm L}=(u_0,\Lambda)\) and radius
\(e\Lambda\), and only its part in \(v>0\) is physical.  Independent
upper-half-plane dilations have been used to give the three Landau
carrier circles the same displayed Euclidean radius; this does not
change \(e\) or the classification.  Consequently, the solid
hypercycle meets \(v=0\) exactly at the ideal endpoints \(I_\pm\);
the lower dashed arc lies outside \(H^2\). Only the attractive Kepler sector \(\gamma>0\) is displayed; the
repulsive sector \(\gamma<0\), represented by the corresponding
minor hodograph and Landau arcs, is described in the text.}
\label{fig:hodograph-Landau}
\end{figure}

Figure~\ref{fig:hodograph-Landau} displays the common geometric
classification.  For \(e>1\), the physical Kepler hodograph is the
solid major arc ending at the points \(P_\pm\).  The rays from the
momentum origin \(O_p\) to these endpoints are tangent to the
hodograph circle and determine the two asymptotic velocity
directions.  On the Landau side, the physical hypercycle is precisely
the part of the Euclidean carrier circle lying in \(v>0\); it meets
the ideal boundary \(v=0\) at the two endpoints \(I_\pm\).  The dashed
portions therefore belong to the corresponding Euclidean carrier
circles but not to the physical Kepler or Landau trajectories.

The figure shows only the attractive sector \(\gamma>0\), for which
the positive homothety factor in \eqref{scaled-direct-map} implies
\(\mathcal B/T>0\).
 The repulsive Kepler sector
\(\gamma<0\) is necessarily an open sector with
\(E_{\rm K}>0\) and \(e>1\).
In this case
\(\mathcal B/T<0\), and the center of the corresponding Landau
carrier circle lies below the ideal boundary \(v=0\).  Only its upper
minor arc belongs to \(H^2\).  Correspondingly, the physical
repulsive Kepler hodograph is the minor arc of its circular carrier
between the two tangency points determined by the momentum origin,
rather than the major arc occurring in the attractive scattering
case.  The repulsive correspondence therefore produces no additional
geometric class: it gives a hypercycle with the opposite signed
geodesic curvature.  For this reason it is not displayed separately
in Fig.~\ref{fig:hodograph-Landau}.

\subsection{Binet-adapted hyperbolic geometry}

The hyperbolic geometry suggested by the carrier circles can be
constructed directly on the regular reduced Kepler phase space.  Fix
\(\ell\neq0\) and introduce
\begin{equation}
 Q=\frac1r>0,
 \qquad
 P=\frac{p_r}{\ell}=-Q'.
 \label{Binet-phase-variables}
\end{equation}
The physical reduced symplectic form then becomes
\begin{equation}
 \omega_\ell=dr\wedge dp_r
 =\ell\,\frac{dP\wedge dQ}{Q^2}.
 \label{Kepler-reduced-hyperbolic-area}
\end{equation}
After normalization by the fixed signed value of \(\ell\),
\begin{equation}
 \Omega_{\rm Bin}:=\frac{\omega_\ell}{\ell}
 =\frac{dP\wedge dQ}{Q^2},
 \label{Binet-normalized-area}
\end{equation}
which is the invariant area form of the Poincar\'e upper half-plane.
An area form alone does not determine a metric, but the Binet
evolution supplies the missing complex structure.  Indeed, with
\(c_{\rm K}=\gamma/\ell^2\), Eq.~\eqref{Kepler-Binet-direct} is
equivalent to
\begin{equation}
 P'=Q-c_{\rm K},
 \qquad
 (Q-c_{\rm K})'=-P.
 \label{Binet-first-order-complex}
\end{equation}
Introducing the shifted Binet vector
\begin{equation}
 \boldsymbol{\xi}
 =
 \begin{pmatrix}
 P\\[1mm]
 Q-c_{\rm K}
 \end{pmatrix},
 \qquad
 c_{\rm K}=\frac{\gamma}{\ell^2},
\end{equation}
the first-order Binet system
\eqref{Binet-first-order-complex} can be written as
\begin{equation}
 \frac{d\boldsymbol{\xi}}{d\alpha}
 =
 A_{\rm Bin}\boldsymbol{\xi},
 \qquad
 A_{\rm Bin}
 =
 \begin{pmatrix}
 0&1\\
 -1&0
 \end{pmatrix},
 \qquad
 A_{\rm Bin}^2=- I.
 \label{Binet-evolution-generator}
\end{equation}
Thus \(A_{\rm Bin}\) is the infinitesimal generator of the angular
Binet evolution, while
\(\exp[(\alpha-\alpha_0)A_{\rm Bin}]\) is the corresponding finite
evolution operator.  The relation \(A_{\rm Bin}^2=- I\)
determines an affine complex structure up to an overall sign.  The
sign compatible with the orientation of
\(\Omega_{\rm Bin}\) and with the convention
\[
 g_{\rm Bin}(X,Y)=\Omega_{\rm Bin}(X,JY)
\]
is
\begin{equation}
 J=-A_{\rm Bin},
 \qquad
 J\partial_P=\partial_Q,
 \qquad
 J\partial_Q=-\partial_P.
 \label{Binet-complex-structure}
\end{equation}
The compatible metric is therefore
\begin{equation}
 ds_{\rm Bin}^2
 =
 \frac{dP^2+dQ^2}{Q^2}.
 \label{Binet-hyperbolic-metric}
\end{equation}
This is the standard Poincar\'e upper-half-plane metric.  Its Gaussian
curvature is
\begin{equation}
 K_{\rm Bin}=-1,
\end{equation}
or, equivalently, its two-dimensional scalar curvature is
\(R_{\rm Bin}=-2\).
Since \(J\) is integrable and
\(d\Omega_{\rm Bin}=0\), the triple
\(\left(g_{\rm Bin},J,\Omega_{\rm Bin}\right)\) defines the standard
K\"ahler structure of the Poincar\'e upper
half-plane.  Its K\"ahler character is, of course, well known.  The
nontrivial Kepler-side statement is that the K\"ahler form
\(\Omega_{\rm Bin}\) coincides with the normalized physical reduced
symplectic form \(\omega_\ell/\ell\), while the compatible complex
structure is selected by the angular Binet evolution.  Thus this
hyperbolic K\"ahler geometry is constructed entirely from the regular
Binet-reduced Kepler dynamics; the Landau problem is not needed to
define it.

The energy integral is the Euclidean circle
\begin{equation}
 P^2+\left(Q-\frac{\gamma}{\ell^2}\right)^2
 =\frac{\gamma^2}{\ell^4}+\frac{2E_{\rm K}}{\ell^2}.
 \label{Binet-energy-circle-hyperbolic}
\end{equation}
For \(\gamma\neq0\), its center and radius satisfy
\begin{equation}
 c_{\rm K}=\frac{\gamma}{\ell^2},
 \qquad
 R_{\rm K}=\frac{|\gamma|}{\ell^2}e,
 \qquad
 |k_g|=\frac{|c_{\rm K}|}{R_{\rm K}}=\frac1e.
 \label{Binet-geodesic-curvature}
\end{equation}
Thus attractive Kepler ellipses, parabolas and hyperbolas become,
respectively, hyperbolic circles, horocycles and hypercycles of the
single metric \eqref{Binet-hyperbolic-metric}.  With the orientation
convention adopted above, the sign of \(\gamma\) fixes the sign of the
geodesic curvature \(k_g\); reversing the traversal orientation reverses
the sign of \(k_g\).  In particular, the repulsive physical sector
\(\gamma<0\), \(E_{\rm K}>0\) gives oppositely oriented hypercycles.
For \(\gamma=0\), the physical Binet curve is a Euclidean semicircle
centered on \(Q=0\); it is a geodesic of
\eqref{Binet-hyperbolic-metric}.  The circular case \(e=0\) is the
degenerate fixed point discussed below.

Recall that
\begin{equation}
 U=u-u_0,
 \qquad
 u_0=\frac{J_0}{T},
\end{equation}
where \(u_0\) is constant on the selected Landau orbit.  Hence
\(dU=du\), and the translation \(u\mapsto U=u-u_0\) is an isometry
of the Poincar\'e upper half-plane.  In terms of this shifted
coordinate, the direct Landau map \eqref{scaled-direct-map} is simply
\begin{equation}
 U=aP,
 \qquad
 v=aQ.
 \label{Binet-Landau-isometry}
\end{equation}
Consequently,
\begin{equation}
 \frac{dU^2+dv^2}{v^2}
 =
 \frac{dP^2+dQ^2}{Q^2},
 \label{Binet-Landau-metric-identity}
\end{equation}
so the scale \(a\) cancels and the direct orbit map identifies the
Binet-adapted Kepler metric with the Poincar\'e
configuration-space metric of the hyperbolic Landau problem.

The hodograph construction gives an orbitwise momentum-space
realization of the same geometry.  Define
\begin{equation}
 \eta
 =
 p_\parallel+\frac{\rho(1-e^2)}{e},
 \qquad
 \mu
 :=
 \frac{\Lambda e}{\rho}>0,
 \label{adapted-hodograph-height}
\end{equation}
so that
\begin{equation}
 (U,v)
 =
 \mu\left(-p_\perp,\eta\right).
\end{equation}
On \(\eta>0\),
\begin{equation}
 ds^2_{\rm hod}
 =
 \frac{dp_\perp^2+d\eta^2}{\eta^2},
 \qquad
 p_\perp^2+
 \left(\eta-\frac{\rho}{e}\right)^2
 =
 \rho^2,
 \qquad
 |k_g|=\frac1e.
 \label{hodograph-hyperbolic-metric}
\end{equation}
Unlike the Binet metric, this realization is orbit-dependent: its
shift, ideal line and longitudinal direction depend on the orbit
constants.  It is therefore not a universal metric on the entire
Kepler momentum plane.  This also distinguishes it from the fixed
hyperbolic velocity space of special relativity
\cite{FockRelativityVelocity}.  The Binet construction complements
the familiar regularizations in which Kepler configuration-space
dynamics at negative, zero and positive energy is related to geodesic
motion on spherical, Euclidean and hyperbolic spaces
\cite{MoserRegularization,LigonSchaaf,HeckmanDeLaat}.

\subsection{Circular, horocyclic and radial limiting cases}

At \(e=0\), the Kepler orbit is circular and its hodograph is a circle
centered at the momentum origin.  The reduced Binet motion itself is
the fixed point
\begin{equation}
 q=\frac{\gamma}{\ell^2},
 \qquad q'=0.
\end{equation}
Under Eq.~\eqref{scaled-direct-map} it maps to
\begin{equation}
 v=\Lambda=\frac{\mathcal B}{T},
 \qquad p_v=0,
 \qquad {\cal H}_{\mathcal B}=0,
 \label{circular-fixed-point}
\end{equation}
which is a stationary Landau point.  There is no contradiction with
the continuing angular Kepler motion: reduction by the Kepler angle
has collapsed that motion to the radial fixed point.  Correspondingly,
the affine hodograph map \eqref{hodograph-Landau-affine} degenerates at
\(e=0\).

For \(e=1\) and \(T\neq0\), the carrier circle is tangent to
\(v=0\) at a finite boundary point.  A horocycle centered at infinity
is the horizontal line \(v=v_*\).  It is obtained in the distinct
parabolic sector
\begin{equation}
 T=p_u=0,
 \qquad
 p_v=0,
 \qquad
 {\cal H}_{\mathcal B}=\mathcal B^2,
 \qquad
 \dot v=0,
 \qquad
 \dot u=-2\mathcal Bv_*.
 \label{horizontal-horocycle-sector}
\end{equation}
Thus \(p_u=0\) does not imply \(\dot u=0\): the canonical momentum
vanishes, whereas the gauge-covariant horizontal momentum is
\[
 p_u-\frac{\mathcal B}{v_*}
 =
 -\frac{\mathcal B}{v_*}.
\]
Finite-center and horizontal horocycles are related by a M\"obius
isometry of \(H^2\), but the horocyclic reduction adapted to the
ideal point at infinity places them in distinct momentum sectors:
\(T\neq0\) for a finite-center carrier and \(T=0\) for a horizontal
horocycle.

The Kepler sector \(\ell=0\) is likewise singular for the direct map:
the polar angle ceases to be an evolution parameter and attractive
radial motion reaches the collision.  In the normalization \(a=1\),
Eq.~\eqref{scaled-direct-map} imposes
\begin{equation}
 T=\frac{\mathcal B}{\gamma}\ell^2,
 \label{radial-horizontal-limit}
\end{equation}
so the formal limit \(\ell\to0\) reaches the horizontal sector
\(T=0\).  Simultaneously, the hodograph radius
\(\rho=|\gamma/\ell|\) diverges and the circle tends to a straight-line
limit.  This identifies the relevant boundary of the regular
correspondence, but it is not a nonsingular one-to-one map of the
Kepler collision orbit to a Landau orbit.

\subsection{Relation to the actual and virtual Morse bridges}

At the level of constants, the arbitrary scale \(a\) allows the direct
construction to be restricted to the same parameter sector as the
original Kepler--Morse--Landau dictionary.  The coordinate
identifications are complementary: the Morse evolution law compares
the Landau height with the Kepler radius, whereas the carrier-circle
map \eqref{scaled-direct-map} compares it with the inverse radius.
Write
\(h={\cal H}_{\rm K}=2E_{\rm K}\).  On the genuine bound-state Morse
branch, \(h<0\), one has
\begin{equation}
 T=C=\sqrt{-h},
 \qquad
 \mathcal B=\frac{\gamma}{C},
 \qquad
 a=\frac{\ell^2}{-h},
 \qquad
 {\cal H}_{\mathcal B}-\mathcal B^2=-\ell^2.
 \label{actual-Morse-direct}
\end{equation}
For \(h>0\), the direct real classical relation can instead be written
as
\begin{equation}
 T=\sqrt h,
 \qquad
 \mathcal B=\frac{\gamma}{T},
 \qquad
 a=\frac{\ell^2}{h},
 \qquad
 {\cal H}_{\mathcal B}-\mathcal B^2=+\ell^2.
 \label{virtual-Morse-direct}
\end{equation}
The latter may be viewed as a ``virtual Morse bridge'': the common
Binet or Whittaker structure remains useful, but the intermediate
Kepler--Morse substitution would require analytic continuation of its
scale.  It is not an ordinary real bound Morse Hamiltonian and, by
itself, does not establish a unitary equivalence of the positive-energy
Kepler and Landau quantum problems.  Any such quantum correspondence
would require a separately constructed intertwining operator together
with its measure, domains and scattering boundary conditions.

\section{Direct reduced quantum Kepler--Landau correspondence}
\label{sec:direct-quantum-KL}

The two quantum Liouville transformations can be composed so that the
intermediate Morse wavefunction disappears from the final formula.
The result is an exact direct relation between the separated Kepler
radial equation and the horocyclically reduced Landau equation on the
attractive negative-energy branch.  
Here ``direct'' means that the intermediate Morse representation has
been eliminated from the final relation: the composed transformation
maps separated Kepler radial wavefunctions directly to
horocyclically reduced Landau wavefunctions.  This construction does
not amount to a quantization of the orbitwise classical map of
Sec.~\ref{sec:direct-KL}.  Since that map is not by itself a canonical
transformation between the unreduced phase spaces, its quantization
would first require its extension to a canonical relation in a
parametrized and possibly oxidized phase space, followed by the
construction of the corresponding quantum intertwiner.

\subsection{Composed wavefunction map and operator-pencil identity}

Write the radial Kepler equation \eqref{radial-Kepler-quantum} as
\begin{equation}
 \widehat h_{{\rm K},\ell}u=h u,
 \qquad
 \widehat h_{{\rm K},\ell}
 =-\frac{d^2}{dr^2}
 +\frac{\ell^2-\frac14}{r^2}
 -\frac{2\gamma}{r},
 \qquad h=2E_{\rm K}<0.
 \label{direct-quantum-Kepler-operator}
\end{equation}
On the common parameter locus, put
\begin{equation}
 C=\sqrt{-h}>0,
 \qquad
 \mathcal B=\frac{\gamma}{C},
 \qquad
 \mathcal E_{\rm L}
 =\mathcal B^2+\frac14-\ell^2.
 \label{direct-quantum-parameter-locus}
\end{equation}
Before the Landau half-density transformation, the reduced operator
acting in the natural measure \(e^X dX\) is
\begin{equation}
 \widehat{\mathcal H}^{\rm red}_{\mathcal B,C}
 =-\frac{d^2}{dX^2}-\frac{d}{dX}
 +\left(Ce^{-X}-\mathcal B\right)^2.
 \label{direct-quantum-Landau-operator}
\end{equation}

Let \({\cal L}_{\rm K}\) denote the Kepler Liouville map
\eqref{Liouville-r-X} and \({\cal L}_{\rm L}\) the Landau
half-density map used in deriving \eqref{quantum-Landau-Morse}.
Their composition is
\begin{equation}
 \mathcal T^{\rm red}_{\rm K-L}
 =\mathcal L_{\rm L}^{-1}\mathcal L_{\rm K},
 \qquad
 (\mathcal T^{\rm red}_{\rm K-L}u)(X)
 =u(e^{-X}).
 \label{direct-quantum-wavefunction-map}
\end{equation}
Thus, in the upper-half-plane coordinate \(v=e^{-X}\), the reduced
Landau wavefunction is simply
\begin{equation}
 \Phi_C(v)=u_\ell(v),
 \label{direct-quantum-Phi-u}
\end{equation}
up to normalization.  The simplicity of this formula results from
the cancellation of the two half-density factors; the nontrivial
content is the metamorphosis of parameters in
\eqref{direct-quantum-parameter-locus}.

Using
\begin{equation}
 \left(\frac{d^2}{dX^2}+\frac{d}{dX}\right)
 \mathcal T^{\rm red}_{\rm K-L}u
 =e^{-2X}\mathcal T^{\rm red}_{\rm K-L}\frac{d^2u}{dr^2},
 \label{direct-quantum-derivative-identity}
\end{equation}
one obtains the exact operator identity
\begin{equation}
 \left(
 \widehat{\mathcal H}^{\rm red}_{\mathcal B,C}
 -\mathcal E_{\rm L}
 \right)\mathcal T^{\rm red}_{\rm K-L}
 =e^{-2X}\mathcal T^{\rm red}_{\rm K-L}
 \left(\widehat h_{{\rm K},\ell}-h\right).
 \label{direct-quantum-operator-pencil}
\end{equation}
Consequently, \(\mathcal T^{\rm red}_{\rm K-L}\) maps the kernel of
the Kepler radial pencil to the kernel of the reduced Landau pencil.
The multiplication factor \(e^{-2X}\) is essential.  In the
nonunitary conformal bridges discussed, for example, in
Refs.~\cite{InzPlyWipf,AlcalaPlyushchay}, two fixed operators are
related by a similarity transformation of the schematic form
\begin{equation}
 \widehat{\mathcal O}_2
 =
 S\widehat{\mathcal O}_1S^{-1}.
\end{equation}
Equation~\eqref{direct-quantum-operator-pencil} cannot be written in
this form.  It is instead a weighted intertwining relation between
parameter-dependent differential-operator pencils: it maps the
kernel of the Kepler radial pencil to the kernel of the reduced
Landau pencil on the matched parameter locus.  This is the direct
operator realization of the St\"ackel transformation and
coupling-constant metamorphosis.

\subsection{Bound states, measures and the threshold endpoint}

For a Kepler bound state, the radial normal-form wavefunction is, up
to normalization,
\begin{equation}
 u_{n_r,\ell}(r)
 \propto
 (2Cr)^{|\ell|+\frac12}e^{-Cr}
 L_{n_r}^{2|\ell|}(2Cr).
 \label{direct-quantum-Kepler-wavefunction}
\end{equation}
Equation~\eqref{direct-quantum-Phi-u} therefore gives the reduced
Landau state by replacing \(r\) with \(v\).  The radial quantum number
is preserved, \(n_{\rm L}=n_r\), while
\begin{equation}
 |\ell|=\mathcal B-n_{\rm L}-\frac12,
 \label{direct-quantum-ell-threshold-distance}
\end{equation}
measures the distance of the Landau level from the continuum
threshold.

The map is not unitary in the physical reduced measures.  Indeed,
\begin{equation}
 \|u\|_{\rm K}^2
 =\int_0^\infty |u(r)|^2dr,
 \qquad
 \|\mathcal T^{\rm red}_{\rm K-L}u\|_{\rm L}^2
 =\int_0^\infty\frac{|u(v)|^2}{v^2}dv.
 \label{direct-quantum-norms}
\end{equation}
Near the ideal boundary, \(\Phi_C(v)\sim
v^{|\ell|+1/2}\), and hence the Landau norm density behaves as
\(v^{2|\ell|-1}\).  It is integrable for \(|\ell|>0\), whereas
\(\ell=0\) produces the logarithmic divergence \(\int_0dv/v\).
Thus the zero-angular-momentum Kepler bound state is mapped to the
bounded, non-normalizable Landau threshold state identified in
Sec.~\ref{sec:quantum-spectrum}.  The regular bridge acts on
\(|\ell|>0\) bound sectors and reaches \(\ell=0\) only in the sense of
generalized states.

\subsection{A fixed-shell grading intertwiner}

The shell-to-tower correspondence of Sec.~\ref{sec:quantum-spectrum}
can now be stated as an operator relation.  Fix
\begin{equation}
 \mathcal B_N=N+\frac12,
 \qquad
 C_N=\frac{\gamma}{\mathcal B_N},
 \qquad
 E_{{\rm K},N}=-\frac{\gamma^2}{2\mathcal B_N^2}.
 \label{direct-quantum-fixed-shell-data}
\end{equation}
For \(N=0\) the shell consists only of the \(\ell=0\) threshold
endpoint.  In the following take \(N\geq1\).  On this Kepler shell,
\(n_r+|\ell|=N\).  Let
\begin{equation}
 \widehat L=-i\frac{\partial}{\partial\varphi},
 \qquad
 \widehat G_{{\rm K},N}
 =\mathcal B_N^2+\frac14-\widehat L^2,
 \label{direct-quantum-shell-grading}
\end{equation}
be the shell grading operator.  For a fixed orientation
\(\sigma=\operatorname{sgn}\ell\), define
\(\mathcal S_N^{(\sigma)}\) by applying
\(\mathcal T^{\rm red}_{\rm K-L}\) to each separated radial component
and retaining \(\sigma\) as a target label.  Explicitly, for
\begin{equation}
 \ell=\sigma(N-n),
 \qquad n=0,1,\ldots,N-1,
 \label{direct-quantum-shell-labels}
\end{equation}
one sets
\begin{equation}
 \mathcal S_N^{(\sigma)}
 \left[
 \frac{e^{i\ell\varphi}}{\sqrt{2\pi r}}u_{n,\ell}(r)
 \right]
 =
 \left(\mathcal T^{\rm red}_{\rm K-L}u_{n,\ell}\right)(X)
 \otimes|\sigma\rangle .
 \label{direct-quantum-shell-map}
\end{equation}
Equations
\eqref{Landau-discrete-levels} and
\eqref{direct-quantum-parameter-locus} then imply
\begin{equation}
 \widehat{\mathcal H}^{\rm red}_{\mathcal B_N,C_N}
 \mathcal S_N^{(\sigma)}
 =\mathcal S_N^{(\sigma)}\widehat G_{{\rm K},N}.
 \label{direct-quantum-shell-intertwiner}
\end{equation}
The relation extends distributionally to the \(n=N\), \(\ell=0\)
threshold state.  Hence one fixed \((2N+1)\)-dimensional Kepler shell
is reorganized into the complete finite reduced Landau tower plus its
threshold endpoint.  The reduced Landau equation depends only on
\(\ell^2\) and suppresses the two orientations
\(\ell=\pm|\ell|\); these must be retained as an additional label in
any one-to-one completion.

Equation~\eqref{direct-quantum-shell-intertwiner} is stronger than a
matching of eigenvalue formulas.  On the fixed shell the Kepler
Hamiltonian selects the finite module, while its angular-momentum
grading is transformed into the reduced Landau Hamiltonian.  It does
not, however, identify the complete Kepler Hamiltonian with a single
Landau Hamiltonian: different Kepler shells determine different
values of \(C\) and \(\mathcal B\).

\subsection{Scope of the direct quantum relation}

The exact construction in this section is physical and real for
\(E_{\rm K}<0\), where \(C=\sqrt{-2E_{\rm K}}\).  For
\(E_{\rm K}>0\), the same Morse composition gives \(C\) and
\(\mathcal B=\gamma/C\) imaginary; at \(E_{\rm K}=0\) it gives a
singular limit.  These analytic continuations do not describe a real
Landau momentum and magnetic field.  
By contrast, the direct classical map of
Sec.~\ref{sec:direct-KL} relates the real elliptic, parabolic and
hyperbolic sectors.  In particular, its hodograph--Landau component
maps Kepler momentum variables to Landau configuration variables.
This momentum--coordinate exchange indicates that a quantum lift of
the correspondence should involve a change of polarization.  After
the orbitwise map is extended to a canonical relation in a
parametrized and possibly oxidized phase space, its microlocal
quantization should therefore be represented by a
polarization-changing Fourier-integral operator.  Such an operator
should intertwine the corresponding parametrized Schr\"odinger
constraint equations, with the appropriate domains, measures and
scattering boundary conditions.  This unreduced quantum completion
is not constructed here; Eqs.~\eqref{direct-quantum-operator-pencil}
and \eqref{direct-quantum-shell-intertwiner} provide exact
consistency conditions that it should satisfy.

\section{Discussion and outlook}
The results obtained above have three logically distinct levels.  The
two Morse reductions are exact statements about separated classical
and quantum equations.  The direct map
\eqref{scaled-direct-map}--\eqref{direct-time-map} is an exact
correspondence of orbit data and reparametrized classical evolution.
The composed quantum relations
\eqref{direct-quantum-operator-pencil} and
\eqref{direct-quantum-shell-intertwiner} are exact reduced
operator-pencil statements.  A fourth, stronger issue is whether a
single global canonical transformation mapping the complete
unreduced Kepler dynamics to the complete unreduced Landau dynamics
can be constructed; such a transformation is not established here.
The reduced relations suggest that
such a completion, if it exists, should be sought in a parametrized
and possibly oxidized phase space.  At the quantum level, a unitary
equivalence of the two fixed complete Hamiltonians is neither claimed
nor expected, since their spectra and spectral multiplicities are
different.

The bridge relates flat electric dynamics to curved magnetic dynamics.
The planar Kepler problem is controlled by the Coulomb coupling
\(\gamma\), while the hyperbolic Landau problem on \(H^2\) is
controlled by the magnetic field \(\mathcal B\).  In the common Morse
normalization, the Kepler--Morse dictionary gives
\[
 \gamma=C\lambda,
 \qquad
 C^2=-2E_{\rm K}
\]
on the physical negative-energy branch, whereas the general signed
horocyclic Landau reduction gives
\[
 C=|p_y|,
 \qquad
 \lambda=\mathcal B\,\operatorname{sgn}(p_y).
\]
The two dictionaries therefore combine into the signed cross-bridge
relation
\begin{equation}
 \gamma=\mathcal Bp_y.
 \label{KML-signed-cross-bridge}
\end{equation}
The detailed real quantum bridge developed above uses the attractive
orientation
\[
 \mathcal Bp_y>0;
\]
with the convention \(\mathcal B>0\), this gives \(p_y=C>0\) and
\(\gamma>0\).  The opposite-sign case
\(\mathcal Bp_y<0\) corresponds to the repulsive reduced sector
described in the classical discussion, but it has no
bound-state quantum tower.  Thus the Coulomb coupling is represented
on the Landau side by the signed product of the magnetic field and
the conserved horocyclic momentum.  This suggests an
electric--magnetic metamorphosis rather than a strict duality.

One should also keep in mind the different spectral status of \(C\)
on the two sides of the bridge.  On the Landau side, \(p_y\) is a
continuous momentum associated with translations along horocycles,
so \(C=|p_y|\) is the corresponding continuous positive Morse scale.
On the Kepler side, by contrast, if the attractive Coulomb coupling
\(\gamma>0\) is fixed and bound-state boundary conditions are imposed,
the same parameter is discretized according to
\[
 C=\frac{\gamma}{n_r+|\ell|+\frac12}.
\]
Thus the bridge first relates reduced differential equations and their
parameters; the distinct Hilbert-space quantizations of the parent
systems then select different subsets or interpretations of the same
Morse spectral data.  The operator-pencil identity
\eqref{direct-quantum-operator-pencil} makes this reduced quantum
relation explicit.  At \(\mathcal B_N=N+1/2\), the stronger shell
relation \eqref{direct-quantum-shell-intertwiner} shows that the
Kepler angular-momentum grading becomes the finite reduced Landau
energy grading, with the \(\ell=0\) endpoint at threshold.

The first distinguished family is
\(A_{\rm M}=N\), or \(\mathcal B=N+\frac12\). At these values the
would-be highest discrete level reaches the continuum edge, the Morse
problem has a bounded threshold resonance, and the reflection amplitude
contains a finite Blaschke-type factor. 
A second special value is the boundary point of the selected
positive-field branch,
\[
 A_{\rm M}=-\frac12,
 \qquad
 \lambda=\mathcal B=0,
 \qquad
 \gamma=0 .
\]
Then the Morse potential degenerates into the pure exponential, or
Liouville, wall
\[
 V_{\rm M}(X)=C^2e^{-2X}.
\]
On the hyperbolic side this is the zero-field geodesic case, 
reduced at fixed nonzero horocyclic momentum, with
\(C=|p_y|\).
 On the Kepler side it is the
zero-Coulomb limit of the radial problem, leaving only the inverse
square term. Quantum mechanically this gives the de Alfaro--Fubini--
Furlan conformal-mechanics Hamiltonian \cite{DFF,AIPT,PioWal,AchLiv}
\[
 -\frac{d^2}{dr^2}+\frac{\ell^2-\frac14}{r^2},
\]
which is mapped by \(r=e^{-X}\) to the Liouville wall. For \(|\ell|>0\)
 the inverse-square coupling is repulsive, while for
\(\ell=0\) it reaches the critical value \(-1/4\). This borderline
case is the one naturally related to scale anomaly and
Berry--Keating-type discussions of scale-invariant spectral problems \cite{BerKeat,SieRod,BenderBrodyMuller};
the genuinely supercritical regime would correspond to going below
this critical coupling \cite{AlcPlyu}.

The symmetry structures are related in a nontrivial way, but the
magnetic isometry algebra and the Morse potential algebra should not
be identified literally.  For fixed \(\mathcal B\), the Landau
problem has a genuine magnetic \(SL(2,\mathbb R)\) symmetry.  
After fixing
\[
 J_-=p_y=T,
 \qquad
 C=|T|,
\]
in the horocyclic reduction, the remaining generators \(J_0\) and
\(J_+\) do not preserve the reduced sector.
  Consequently, the
magnetic \(SL(2,\mathbb R)\) symmetry of the parent system is not
inherited as an ordinary Noether symmetry of a single fixed Morse
Hamiltonian.

Nevertheless, its classical Casimir relation
\[
 {\cal C}_{\rm L}
 =
 {\cal H}_{\mathcal B}-\mathcal B^2
\]
descends under reduction to the classical Morse Hamiltonian.
Quantum mechanically, the corresponding reduced spectral relation
contains the additional half-density shift \(1/4\), as displayed in
Eq.~\eqref{Morse-Landau-energy-quantum}.  A second,
complementary algebraic structure
is supplied by the Darboux
intertwiners, which shift \(A_{\rm M}\), or equivalently
\(\mathcal B\), by integers and therefore connect different members
of the Morse--Landau family.  Thus the Casimir reduction of the
parent magnetic symmetry and the parameter-shifting Morse potential
algebra encode two distinct but complementary aspects of the bridge.

The planar Kepler problem has the energy-dependent Laplace--Runge--Lenz
symmetry, yielding \(so(3)\), \(e(2)\), or \(so(2,1)\) after fixing
negative, zero, or positive energy \cite{BandItzi1,BandItzi2,BacryRueggSouriau}. The Morse system exchanges energy,
coupling constants, separation constants and Casimirs, making this
mismatch part of the correspondence rather than an obstruction.

There is also a dynamical-conformal aspect behind this structure.
Equation~\eqref{Landau-x-linear-equation} provides a Binet-type
linearization of the Landau motion:
\begin{equation}
 \ddot x
 =
 4\left(
 {\cal H}_{\mathcal B}-\mathcal B^2
 \right)x
 +
 4\mathcal BT,
 \qquad
 x=\frac1v.
\end{equation}
According to the sign of
\({\cal H}_{\mathcal B}-\mathcal B^2\), this is the equation of a
shifted oscillator, a constant-force system, or a shifted inverted
oscillator.  The Morse motion has exactly the same affine-linear
form: with
\[
 q=Ce^{-X},
 \qquad
 y_{\rm M}=\frac1q,
\]
one finds
\begin{equation}
 \ddot y_{\rm M}
 =
 4E_{\rm M}y_{\rm M}+4\lambda.
\end{equation}
Thus the negative-, zero- and positive-energy Morse sectors, and the
corresponding magnetic-circle, horocycle and hypercycle sectors of
the Landau problem, are represented dynamically by the elliptic,
parabolic and hyperbolic constant-coefficient linear equations.
This oscillator--constant-force--inverted-oscillator trichotomy closely
parallels the affine extension of the
free-particle--oscillator--inverted-oscillator conformal triangle
studied in Ref.~\cite{AlcPlyuAffine}.

The Kepler Binet equation,
\begin{equation}
 \frac{d^2Q}{d\varphi^2}+Q
 =
 \frac{\gamma}{\ell^2},
 \qquad
 Q=\frac1r,
 \qquad
 \ell\neq0,
\end{equation}
belongs to the same affine-linear family, but with an important
difference: it remains a shifted oscillator equation for all values
of \(E_{\rm K}\).  The Kepler energy enters through its first integral
and the eccentricity, rather than through the coefficient of the
Binet equation.  The construction of Sec.~7 makes the geometric
content of this distinction precise.  The regular Binet-reduced
Kepler phase space carries the fixed Poincar\'e K\"ahler metric
\eqref{Binet-hyperbolic-metric}, independently of the sign of
\(E_{\rm K}\), while its trajectories have constant geodesic
curvature
\[
 |k_g|=\frac1e.
\]
Consequently, the attractive sectors \(E_{\rm K}<0\),
\(E_{\rm K}=0\), and \(E_{\rm K}>0\) appear in one and the same
hyperbolic plane as hyperbolic circles, horocycles and hypercycles,
respectively; the repulsive sector has the opposite signed geodesic
curvature, and the free case \(\gamma=0\) gives geodesics.  The
original projective observation is therefore sharpened into two
complementary layers: an affine-linear dynamical organization in
different variables and evolution parameters, and an intrinsic
hyperbolic K\"ahler geometry of the Kepler Binet plane.  These layers
are distinct from the magnetic isometry algebra and from the Morse
potential algebra, but together they indicate a broader
dynamical-conformal structure of the Kepler--Morse--Landau bridge.
Related conformal-bridge constructions have been developed for
confinement and asymptotic freedom, rotational systems, cosmic-string
backgrounds, \(\mathcal{PT}\)-symmetric and supersymmetric models,
noncommutative Landau dynamics, and projective
free-particle--oscillator maps
\cite{InzPlyWipf,InzPlyuExotic,InzPlyuCosmic,InzPlyuSuper,
AlcPlyuVort,AlcalaPlyushchay,AlcPlyu}.

This also clarifies the role of oxidation.  In the reduced Landau
picture, the positive Morse scale \(C=|p_y|\) is determined by the
fixed horocyclic momentum; on the selected positive-momentum branch,
this reduces to \(C=p_y>0\).  In the Morse
potential algebra, by contrast, the parameter \(A_{\rm M}\) can be
represented as a momentum conjugate to an auxiliary angle, for
instance on a cylinder or punctured plane, possibly with an
Aharonov--Bohm shift
\cite{AharonovBohm,LaidlawDeWitt,GroscheABHyp,CJP}.
Undoing the reduction means promoting such constants back to dynamical
momenta. A direct unreduced Kepler--Landau relation would also have to
undo the second reduction hidden in the Kepler--Morse map: the Morse
time \(\tau\), proportional to the Kepler polar angle, should be
promoted to a dynamical variable with its conjugate separation
constant. The Landau conic law above gives the complementary statement:
the Landau time itself enters the Kepler-like conic parametrization and
therefore should also be treated in an extended phase-space
formulation.
In this sense the present one-dimensional bridge suggests
a route to a higher-dimensional canonical, or oxidized,
correspondence between planar Kepler and hyperbolic Landau dynamics.
Here ``oxidation'' is used in the Hamiltonian sense of restoring
variables frozen by reduction as dynamical variables.  
This is analogous in spirit, but not identical, to dimensional
oxidation or uplift in supergravity
\cite{CreJulSch,Keurentjes}; for a recent application of dimensional
oxidation in the Banks--Fischler--Shenker--Susskind (BFSS) matrix
model and eleven-dimensional supergravity, see Ref.~\cite{DiaSan}.

More precisely, an extended canonical construction would have to
promote the two times and the parameters \(\gamma\) and
\(\mathcal B\) to coordinates with conjugate momenta.  On the
corresponding parametrized constraint surfaces, the extended
Poincar\'e--Cartan one-forms are
\begin{align}
 \Theta_{\rm K}^{\rm ext}
 &=
 p_r\,dr+\ell\,d\varphi
 -{\cal H}_{\rm K}\,dt_{\rm K}
 +\pi_\gamma\,d\gamma,
 \\
 \Theta_{\rm L}^{\rm ext}
 &=
 p_u\,du+p_v\,dv
 -{\cal H}_{\mathcal B}\,dt_{\rm L}
 +\pi_{\mathcal B}\,d\mathcal B.
\end{align}
A local extended canonical map would have to satisfy, patch by patch,
the Poincar\'e--Cartan condition
\begin{equation}
 \Theta_{\rm K}^{\rm ext}
 =
 \Theta_{\rm L}^{\rm ext}
 +
 d\mathcal G,
 \label{extended-PC-condition}
\end{equation}
where \(\mathcal G\) is a local generating function.
The remaining global questions are concrete: continuation through the
singular strata \(\ell=0\) and \(T=0\); patching Landau charts centered
at different ideal points; compatibility of the periodic Kepler angle
with the noncompact horocyclic coordinate; and, quantum mechanically,
the domains, measures and possible Aharonov--Bohm holonomy of the
unreduced intertwiner.  Equation~\eqref{scaled-direct-map} supplies nontrivial
local data for such a construction, especially through the exchange
of a Kepler momentum component with a Landau coordinate, but it does
not by itself solve these global problems.

The same Morse Hamiltonian is a Whittaker system 
\cite{KostantWhit,GiventalToda,GerKharLebObl}
and appears in
boundary Liouville minisuperspace \cite{FZZ,PonsotTeschnerBoundary,ApresyanSarkissian}. Although no holographic duality is
claimed here, 
the construction lies close to
a broader circle of ideas linking \(AdS_2/CFT_1\),
\(SL(2,\mathbb R)\), Liouville and Whittaker systems,
Schwarzian dynamics, JT gravity and the SYK model
\cite{SY,Kitaev,MSY,Jensen,MTV,Mertens}.

In summary, the Kepler--Morse--Landau bridge gives a concrete
flat--hyperbolic, electric--magnetic and spectral--geometric
metamorphosis,
\[
\text{flat Kepler--Coulomb}
\longleftrightarrow
\text{Morse--Whittaker}
\longleftrightarrow
\text{hyperbolic Landau}.
\]
The direct orbit map of Sec.~\ref{sec:direct-KL} adds a second,
classical layer to the Morse bridge.  One of its essential consequences
is the hyperbolic geometry revealed intrinsically on the regular
Binet-reduced Kepler phase space.  For every fixed
\(\ell\neq0\), the normalized reduced symplectic form and the complex
structure selected by the Binet evolution determine the
Poincar\'e metric
\begin{equation}
 ds_{\rm Bin}^2
 =
 \frac{dP^2+dQ^2}{Q^2},
 \qquad
 Q=\frac1r>0,
 \qquad
 P=\frac{p_r}{\ell}.
 \label{Binet-metric-conclusion}
\end{equation}
The Kepler Binet trajectories have constant geodesic curvature
\begin{equation}
 |k_g|=\frac1e .
\end{equation}
Consequently, for the attractive Kepler problem, the sectors
\(E_{\rm K}<0\), \(E_{\rm K}=0\), and \(E_{\rm K}>0\) are represented
in one and the same hyperbolic plane by hyperbolic circles,
horocycles, and hypercycles, respectively.  The repulsive sector has
the opposite signed geodesic curvature, while for \(\gamma=0\) the
Binet trajectories are geodesics.  On the regular sector of the
direct correspondence, the map
\((U,v)=a(P,Q)\) identifies this Binet metric isometrically with the
configuration-space metric of the hyperbolic Landau problem.
Hamilton's Kepler hodographs and the Landau carrier circles then give
complementary affine realizations of the same eccentricity
classification.

This organization differs essentially from the familiar
energy-dependent regularizations of the Kepler problem
\cite{MoserRegularization,LigonSchaaf,HeckmanDeLaat}.  In the planar
case, the negative-, zero-, and positive-energy sectors are usually
related to geodesic motion on \(S^2\), \(\mathbb R^2\), and \(H^2\),
with dynamical algebras \(so(3)\), \(e(2)\), and
\(so(2,1)\simeq sl(2,\mathbb R)\), respectively.  In the present
Binet construction, by contrast, the ambient geometry remains the
same hyperbolic plane for all three energy sectors; the energy sign is
encoded in the extrinsic geodesic curvature of the orbit rather than
in the curvature of the ambient space.  In this sense, the
hyperbolic Binet metric sharpens the conformal and projective
structure already suggested by the Binet linearization: the earlier
\(SL(2,\mathbb R)\)-type observation is promoted here to a
distinguished metric geometry of the reduced Kepler dynamics.  The
Landau correspondence does not merely impose this geometry from
outside, but identifies it with the natural geometry of the
hyperbolic Landau configuration space.

Section~\ref{sec:direct-quantum-KL} adds an exact reduced quantum
layer.  The radial Kepler and horocyclic Landau operator pencils are
connected by the weighted intertwining relation
\eqref{direct-quantum-operator-pencil}, while on a fixed
negative-energy Kepler shell the angular-momentum grading is
reorganized by \eqref{direct-quantum-shell-intertwiner} into the
finite reduced Landau tower together with its threshold endpoint.
The classical Binet isometry and these exact reduced quantum
relations substantially sharpen the Kepler--Landau correspondence.
They also indicate the problem of extending the correspondence to a
global canonical relation in a parametrized and possibly oxidized
phase space.  Quantum mechanically, the corresponding question is
the construction of an intertwiner between the extended
Schr\"odinger constraint equations, rather than a unitary equivalence
of the fixed Kepler and Landau Hamiltonians, which is excluded by
their different spectra and spectral multiplicities.

\section*{Acknowledgments}
The work was partially supported by the FONDECYT Project 1242046.

\end{document}